\documentclass[a4paper,12pt,preprint,nofootinbib]{revtex4}
\usepackage{amsmath,mathrsfs,amscd,graphicx}
\usepackage{mciteplus}
\usepackage{multirow}
\usepackage{color}
\usepackage{rotating}
\usepackage{slashed}
\usepackage{subfig}
\usepackage{textcomp}
\usepackage[final]{pdfpages}
\usepackage{array}
\usepackage{float}
\usepackage{url}
\usepackage{textcomp}
\usepackage{amsfonts, latexsym, epsfig}
\usepackage{bm}
\usepackage{times}
\usepackage{epsfig}
\usepackage{amssymb}
\usepackage{tikz}
\usepackage{cancel}
\usepackage{hyperref}
\usepackage{verbatim}
\usepackage[normalem]{ulem}
\usepackage{amssymb}
\usepackage{appendix}
\newcommand{\nn}{\nonumber}
\def\ifmath#1{\relax\ifmmode #1\else $#1$\fi}
\def\half{\ifmath{{\textstyle{\frac{1}{2}}}}}

\def\beq{\begin{equation}}
\def\eeq{\end{equation}}
\def\be{\begin{equation}}
\def\ee{\end{equation}}
\def\bea{\begin{eqnarray}}
\def\eea{\end{eqnarray}}
\newcommand{\SM}{\text{SM}}

\newcommand{\MSbar}{\overline{{\rm MS}}}
\def\FCA{\mathcal{F}^{(1)}}
\def\FCB{\mathcal{F}^{(2)}}

\def\kappat{}

\def\gw{g_W}

\def\bt{B^{(t)}}
\def\btp{B^{'(t)}}
\def\ct{C^{(t)}}
\begin{document}

\title{Probing the trilinear Higgs boson self-coupling via single Higgs production at the LHeC}
\author{Ruibo Li~$^{a}$}
\email{bobli@zju.edu.cn}
\author{Xiao-Min Shen~$^{a}$}
\email{xmshen@zju.edu.cn}
\author{Bo-Wen Wang~$^{a}$}
\email{0617626@zju.edu.cn}
\author{Kai Wang~$^{a}$}
\email{wangkai1@zju.edu.cn}
\author{Guohuai Zhu~$^{a}$}
\email{zhugh@zju.edu.cn}
\affiliation{$^a$ Zhejiang Institute of Modern Physics, Department of Physics, Zhejiang University, Hangzhou, Zhejiang 310027, CHINA}
%\preprint{IPMU18-0029} 
\begin{abstract}
	The determination of the Higgs self coupling is one of the key ingredients for understanding  the mechanism behind the electroweak symmetry breaking.  An indirect method  for constraining {the Higgs trilinear self coupling via single Higgs production} at next-to-leading order (NLO) has been proposed in order to avoid the drawbacks of studies with double Higgs production. {In this paper we study the Higgs self interaction through the vector boson fusion (VBF) process} $e^{-} p \to \nu_{e} h j$ {at the future LHeC.} {At NLO level, we compute analytically the scattering amplitudes for relevant processes, in particular those induced by the Higgs self interaction.} {A Monte Carlo simulation and a statistical analysis utilizing the analytic results are then carried out {for Higgs production through VBF and decay to $b\bar{b}$, which} yield for the trilinear Higgs self-coupling rescaling parameter $\kappa_{\lambda}$ the limit [-0.57, 2.98] with $2~\text{ab}^{-1}$ integrated luminosity. If we {assume about 10\% of the signal survives the event selection cuts, and include all the background,} the constraint will be broadened to [-2.11, 4.63]}. 
\end{abstract}

\maketitle
\section{\label{intro}Introduction}
A standard model {(SM)}-like Higgs boson has been discovered by the ATLAS and CMS collaborations at the CERN Large Hadron Collider (LHC) individually~\cite{Aad:2012tfa,Chatrchyan:2012xdj}, which makes a milestone in particle physics.  While it strongly supports the SM mechanism of spontaneous electroweak symmetry breaking (EWSB), by which all fermions and some of the vector bosons acquire their masses, the driven force of EWSB still remains mysterious. To better understand this problem, it is crucial to study the properties of {the} Higgs boson, {e.g., to measure} its mass, spin, CP properties and couplings~\cite{Aad:2015zhl,Khachatryan:2014kca,Aad:2015wra,Khachatryan:2016vau}.  From the second run of the LHC at ${13}~\text{TeV}$, {the} ATLAS collaboration has recently reported the results of their measurements {$\mu_{H\rightarrow\tau\tau}=1.09^{+0.36}_{-0.30}$ and $\mu_{H\rightarrow bb}=1.01^{+0.20}_{-0.19}$, with the integrated luminosities $36.1$ fb$^{-1}$ and $79.8$ fb$^{-1}$, respectively}~\cite{Aaboud:2018pen,Aaboud:2018zhk}. These are significant improvements in Higgs precision physics.

{Equally important is the the measurement of the Higgs self-coupling ($\lambda$) from the scalar potential $V(\Phi)$, which takes a form with the trilinear ($\lambda_{3}^{SM}=\lambda$) and quartic ($\lambda_{4}^{SM}=\lambda/{4}$) self interactions:
\bea
V(\Phi)=-\mu^{2}\Phi^{\dagger}\Phi+\lambda(\Phi^{\dagger}\Phi)^{2} \rightarrow \frac{1}{2}m_{h}^{2}h^{2}+\lambda_{3}\nu h^{3}+\lambda_{4}h^{4},
\eea
where $\Phi$ is the Higgs doublet field and $h$ is the Higgs boson.~In addition to its crucial role in EWSB, the value of $\lambda$ has interesting implications on physics beyond the SM (BSM). For example, in electroweak baryogenesis, a large deviation of $\lambda$ from its SM value has been used to explain the observed cosmic baryon-antibaryon asymmetry~\cite{Noble:2007kk,Trodden:1998ym,Morrissey:2012db}.} 

In contrast to the measurement of Higgs-fermion couplings, the study of {$\lambda$} is in a completely different situation.  
%{After EWSB  the scalar potential takes a form with} the trilinear ($\lambda_{3}^{SM}=\lambda$) and quartic ($\lambda_{4}^{SM}=\lambda/{4}$) self interactions:
%\bea
%V(\Phi)=-\mu^{2}\Phi^{\dagger}\Phi+\lambda(\Phi^{\dagger}\Phi)^{2} \rightarrow \frac{1}{2}m_{h}^{2}h^{2}+\lambda_{3}\nu h^{3}+\lambda_{4}h^{4},
%\eea
%where $\Phi$ is the Higgs doublet field and $h$ is the Higgs boson. 
{At the LHC, double Higgs production as the standard process for determining the Higgs trilinear self coupling suffers from a small production rate and huge QCD backgrounds, and thus leads to large uncertainties even after the Run-II upgrade.} The measurements of the $\gamma\gamma b\bar{b}$ final states by the CMS and ATLAS experiments yield the constraints $-11\lambda_{3}^{SM}<\lambda_{3}<17\lambda_{3}^{SM}$ and $-8.2\lambda_{3}^{SM}<\lambda_{3}<13.2\lambda_{3}^{SM}$, respectively~\cite{Sirunyan:2018iwt,Aaboud:2018ftw}. For the $b\bar{b}b\bar{b}$ production, {the observed upper limit by ATLAS using the non-resonant Higgs pair production data is 13 times the SM value at 95\% C.L.}~\cite{Aaboud:2018knk}.~{Combining the measurements of the different final states, CMS and ATLAS report their limits $-11.8\lambda_{3}^{SM}<\lambda_{3}<18.8\lambda_{3}^{SM}$ and $-5.0\lambda_{3}^{SM}<\lambda_{3}<12.0\lambda_{3}^{SM}$~\cite{Sirunyan:2018two,Aad:2019uzh}.} There are also extensive phenomenological studies on determining the trilinear Higgs self-coupling directly at the LHC {(including the prospect studies for the high-energy and high-luminosity upgrades of the LHC)}~\cite{Baur:2002rb,Baur:2002qd,Baur:2003gp,Moretti:2004wa,Dolan:2012rv,Baglio:2012np,Goertz:2013eka,Frederix:2014hta,Cao:2015oxx,Gouzevitch:2013qca,Behr:2015oqq,Bishara:2016kjn,DiVita:2017eyz,Cepeda:2019klc}, the future electron-positron collider~\cite{Baer:2013cma,Asner:2013psa,DiVita:2017vrr,Maltoni:2018ttu}, 
%the Future Circular Collider~\cite{Barr:2014sga,Chen:2015gva,Chang:2018uwu,Blondel:2018aan}, 
and future high energy hadron colliders~\cite{Yao:2013ika,Barr:2014sga,Azatov:2015oxa,He:2015spf,Chen:2015gva,Contino:2016spe,Banerjee:2018yxy,Chang:2018uwu,Blondel:2018aan,Kim:2018uty,Kim:2018cxf,Kim:2019wns}, {in which strict constraints are obtained} with higher integrated luminosities and energies.~On the other hand, an indirect method is proposed for constraining the Higgs self-coupling via single Higgs production at next-to-leading order (NLO)~\cite{McCullough:2013rea,Shen:2015pha,Degrassi:2016wml,Bizon:2016wgr,Maltoni:2017ims,DiVita:2017vrr}. {The method relies on the account of one-loop electroweak radiative corrections to Higgs-strahlung and vector boson fusion (VBF) processes~\cite{Fleischer:1982af,Denner:1992bc,Denner:2003iy,Belanger:2002ik}, and it has the potential of reaching a superior precision in the} determination of the Higgs self-coupling. {From the single Higgs production measurement,  ATLAS obtained recently the constraint $-3.2\lambda_{3}^{SM}<\lambda_{3}<11.9\lambda_{3}^{SM}$~\cite{ATL-PHYS-PUB-2019-009}. The combination of double and single Higgs production measurements gives $-2.3\lambda_{3}^{SM}<\lambda_{3}<10.3\lambda_{3}^{SM}$~\cite{ATLAS:2019pbo}.}

In view of the large QCD backgrounds interfering with the one-loop electroweak radiative corrections at the hadron-hadron collider, the Large Hadron electron Collider (LHeC) has been proposed as a deep inelastic scattering facility for the precision measurement of parton distributions and Higgs properties. LHeC as a relatively economic proposal  is an upgrade based on the current 7 TeV proton beam of the LHC by adding one electron beam with 60--140 GeV energy~\cite{AbelleiraFernandez:2012cc}, which could be tuned into a ``Higgs factory'' in which Higgs bosons are produced via VBF process.~Thanks to the forward detector and reduction of QCD backgrounds in the $e$-$p$ collider, the bottom Yukawa and trilinear Higgs self couplings could be measured precisely~\cite{Han:2009pe,Kumar:2015kca,Kumar:2015tua}. Therefore, we expect the LHeC to be a good facility for studying $\lambda_{3}$ via single Higgs production at NLO {level}.

{{One may try} to constrain $\lambda_{3}$ at the LHeC via Higgs pair production. The process has been studied in Ref.~\cite{Kumar:2015kca}. Combining the pair production cross section at the LHeC with the signal and background selection efficiencies from the work above\footnote{Here we assume no significant change in the selection efficiencies at the LHeC, as compared to those in Ref.~\cite{Kumar:2015kca}. The full analysis of Higgs pair production is out of the scope of this study.}, one can estimate that in order to obtain a 2$\sigma$ signal significance, the integrated luminosity needed would be about 23.6 ab$^{-1}$, much higher than the planned 2 ab$^{-1}$ at the LHeC. Hence we only focus on the study of single Higgs production hereafter. }
   
The paper is organized as follows. {In the next section, we discuss the one-loop contribution to single Higgs production, {in particular that from processes} via the trilinear Higgs self {interaction} and Higgs top quark Yukawa {interaction}, and calculate {their scattering amplitudes analytically.} In section \ref{MC}, we perform a Monte Carlo simulation for single Higgs production at the LHeC, produce the differential and total cross section, and carry out a statistical analysis to {obtain constraints} for $\lambda_{3}$. Finally, we conclude in section \ref{con}. }

\section{\label{calc}The one loop correction to single Higgs production at the LHeC}
\begin{figure}[t]
\centering
{\includegraphics[width=0.2\textwidth]{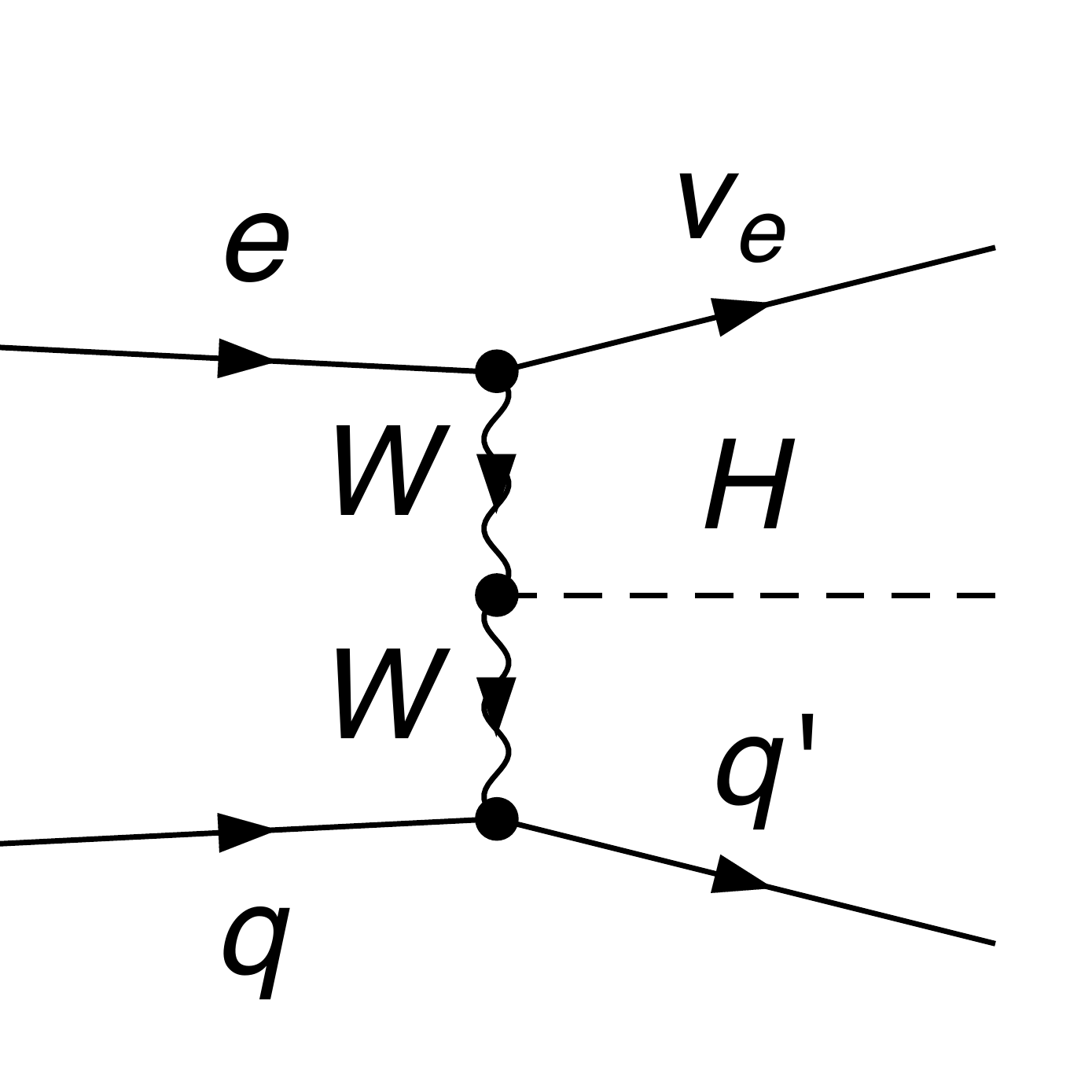}}
	\caption{The Feynman diagram of  single Higgs production {via $W$ boson fusion} at leading order with $q=u,c,\bar{d},\bar{s}$ and $q'=d,s,\bar{u},\bar{c}$ at the LHeC.}
\label{vbf}
\end{figure}

%In the SM, the single Higgs are produced via VBF process $e^{-}p \to \nu_{e} h j$ at the LHeC shown in Fig.\ref{vbf}.
Given the tiny cross section of di-Higgs production~\cite{Kumar:2015kca},
one {could instead constrain} the trilinear Higgs self-coupling $\lambda_{3}$ at the LHeC via the {$\lambda_{3}$ induced loop} corrections to the tree level single Higgs production {process} $e^{-}p \to \nu_{e} h j$ shown in Fig.\ref{vbf}.
We {parameterize} the deviation of possible new physics from SM by a single parameter $\kappa_{\lambda}$:
\begin{equation}
  \label{eq:def_kappa}
  \lambda_{3}^\SM h^3 \to \lambda_3 h^3=\kappa_{\lambda}\lambda_{3}^{SM} h^3,  
\end{equation}
where the physical Higgs field $h$ has a zero vacuum expectation value (VEV), and $\lambda_{3}^{SM} \approx 0.13$ is the {Higgs trilinear self-coupling} in the SM. Other parameters are assumed to be fixed at their SM values~\cite{Degrassi:2016wml,Maltoni:2017ims}.
{Realistic scenarios have been discussed in~\cite{McCullough:2013rea}. For example, the $(|\Phi|^2 - v^2/2)^3/\Lambda^2$ operator, with appropriately tuned parameters, leads to deviation of the $h^3$ coupling, leaving the Higgs mass term, VEV, and other trilinear couplings (such as the couplings of Higgs and Goldstone $hG^+G^-, hG^0G^0$) unchanged.}
\footnote{{This operator, after electroweak symmetry breaking, also leads to deviation of operators with dimension $D \ge 4$,  such as $h^4, h^4G^+G^-$. Our study of the trilinear Higgs coupling (calculation of loop diagrams, expressions of renormalisation constants, wave function corrections), however,  will not be spoild by the possible deviation of these marginal or irrelevant operators.}}

In the following, we shall identify various contributions up to NLO that are relevant for constraining the Higgs trilinear coupling.
	
\subsection{Trilinear Higgs self-coupling: $\lambda_{3}$}
\label{subsec:lambda3}
{
In this section we {consider} only the production of Higgs boson. Note, however, the decay of Higgs boson is also $\lambda_3$-dependent, and has been studied in ~\cite{Degrassi:2016wml}.
}

\begin{figure}[H]
\centering
\subfloat[]{\includegraphics[width=0.2\textwidth]{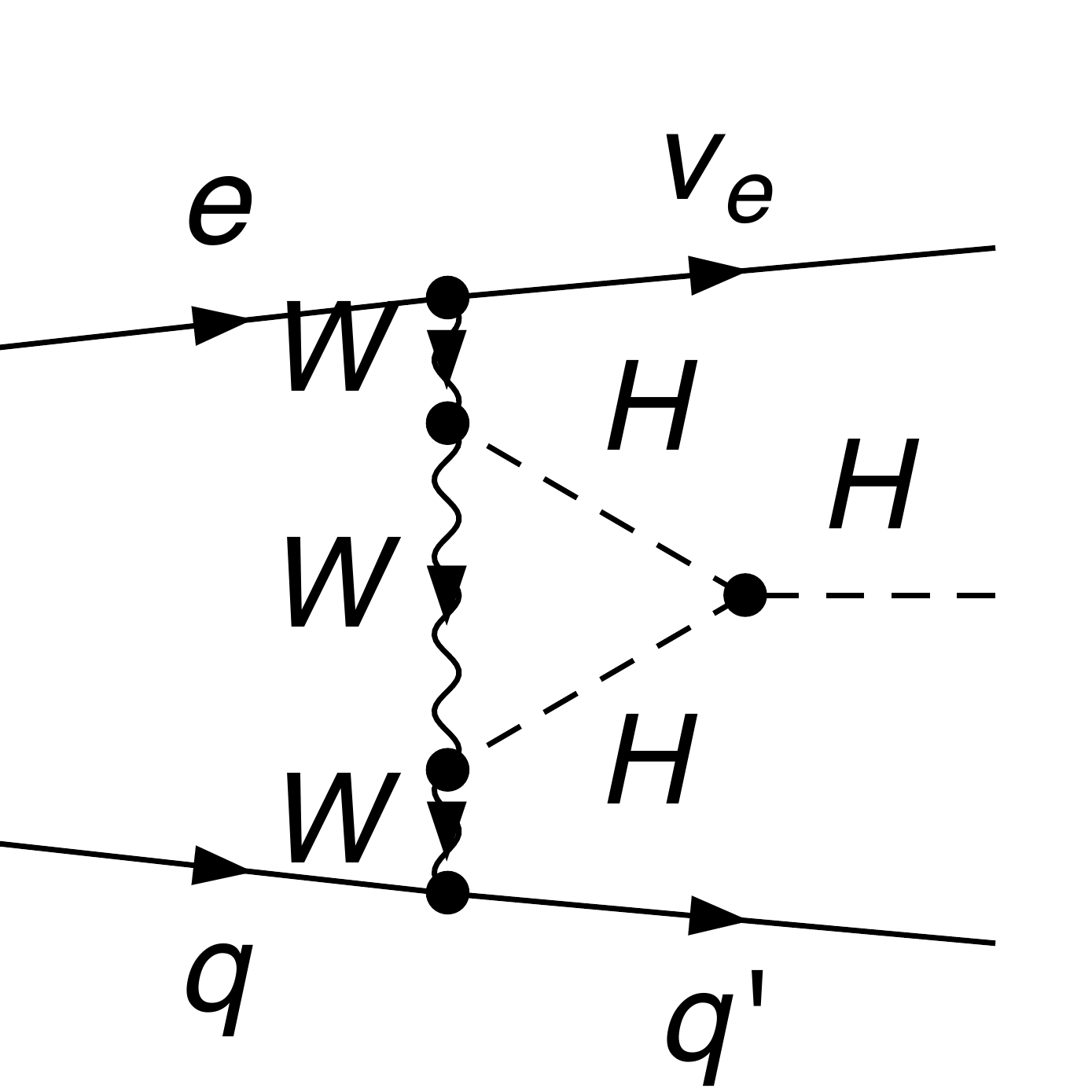}}
\subfloat[]{\includegraphics[width=0.2\textwidth]{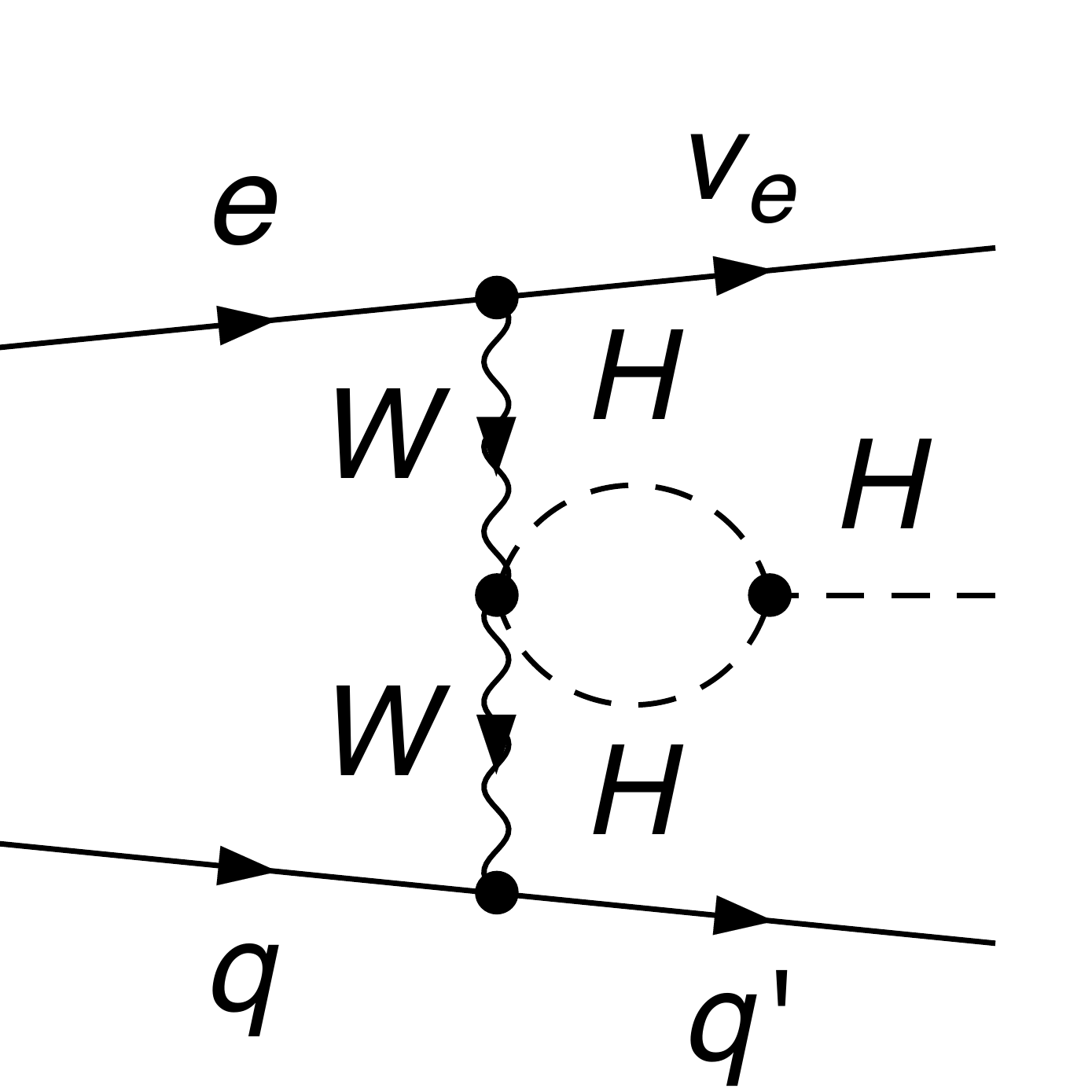}}
\subfloat[]{\includegraphics[width=0.2\textwidth]{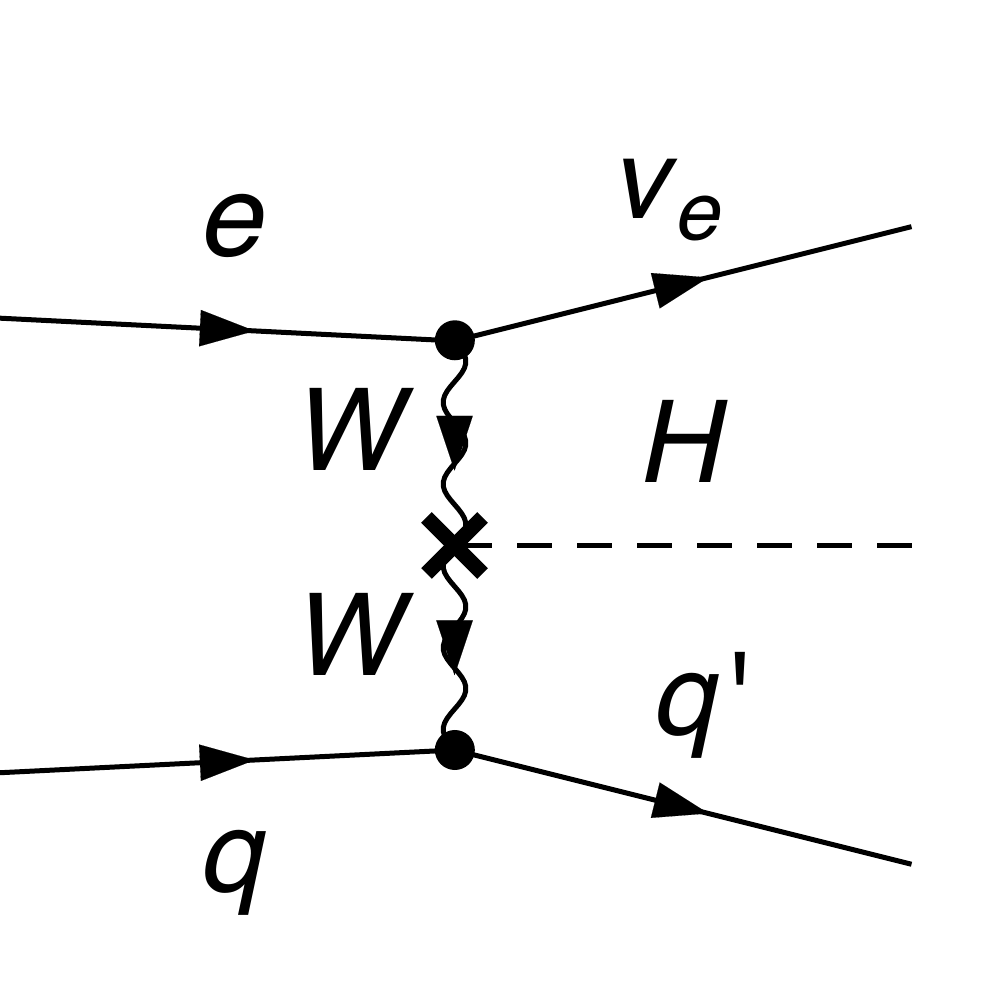}}
\caption{The $\lambda_3$-dependent Feynman diagrams and the corresponding counter term in unitary gauge at one-loop level with $q=u,c,\bar{d},\bar{s}$ and $q'=d,s,\bar{u},\bar{c}$.} 
\label{oneloop}
\end{figure}

{
  For VBF process at the LHeC, the trilinear Higgs self-coupling $\lambda_3$
  enters at one-loop level.
}
{
	The $\lambda_3$-dependent one-loop {corrections to the $HWW$ vertex} are calculated in the unitary gauge.
  The relevant Feynman diagrams are shown in Fig.\ref{oneloop}.
  To check the correctness and gauge invariance of our result,
  we compute the same process in the $R_\xi$ gauge
  (where an additional diagram with Goldstone boson is needed)
  with the {\it FeynArts} and {\it FormCalc} packages~\cite{Hahn:2000kx,Hahn:1998yk}.
  The result is independent of $\xi$ and equals to our result obtained in the unitary gauge.
}

\begin{figure}[H]
\centering
\subfloat[]{\includegraphics[width=0.2\textwidth]{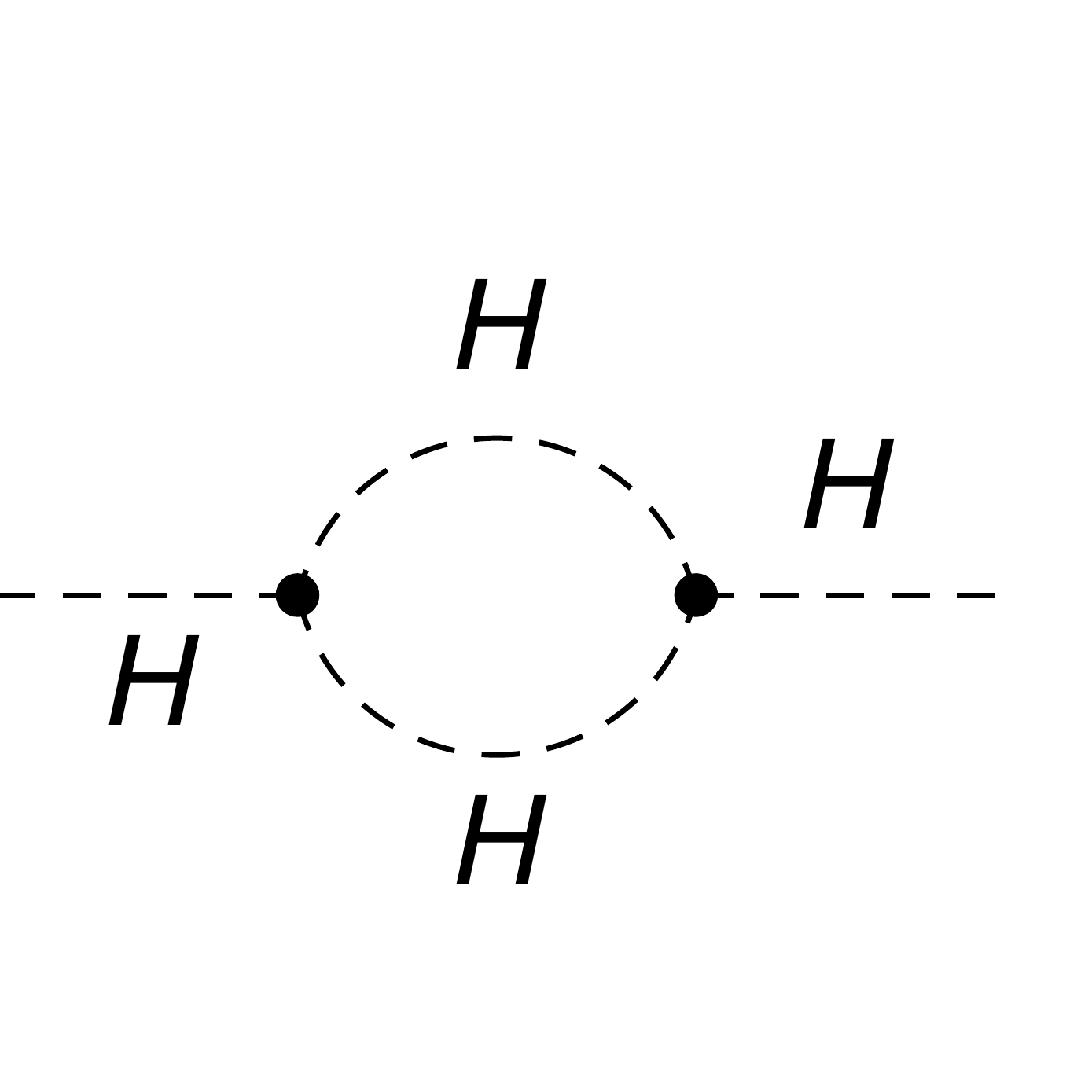}}
\subfloat[]{\includegraphics[width=0.2\textwidth]{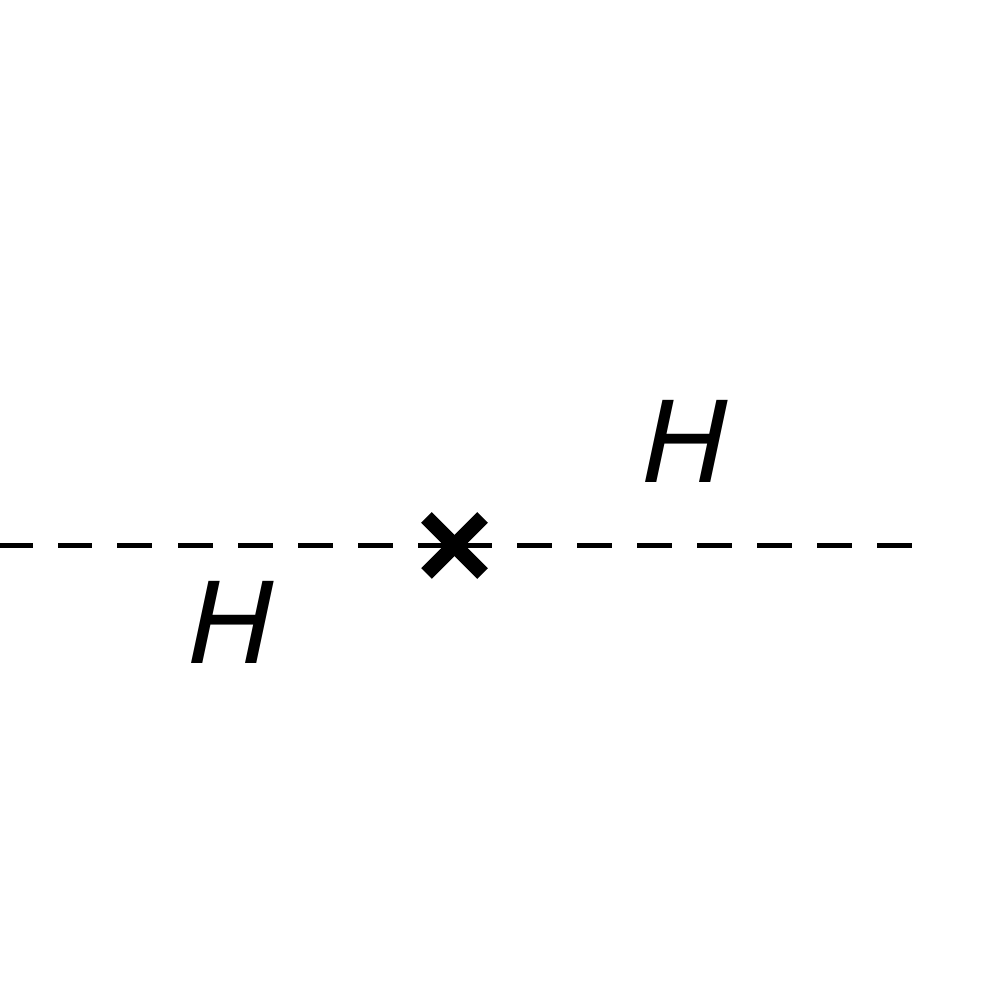}}
	\caption{$\lambda_3$-dependent corrections to the Higgs wave function.}
\label{wave-re}
\end{figure}

{
	The $\lambda_3$-dependence of {Higgs} wave function correction comes from the
  bubble diagram and the counter term diagram, as shown in Fig.\ref{wave-re}.
}

{
  As to the renormalization of the theory, we follow the framework given in Ref.~\cite{Denner:1991kt} and take $\kappa_\lambda, e, M_H, M_W, M_Z $ as input parameters. Note that unlike Ref.~\cite{Denner:1991kt}, which concerns the renormalization of the Standard Model, we introduce an additional input parameter $\kappa_\lambda$. Scrutiny of Ref.~\cite{Denner:1991kt} shows that most of the results there are still valid in our model. Major difference appears in the wave function renomalization constant of Higgs boson, denoted by $\delta Z_H$.~As we will see in section \ref{subsec:result}, the contribution of $\delta Z_H$ {to} the $HWW$ {vertex counter term} cancels its contribution {to} the counter term of Higgs wave function correction. The Higgs field $h$ is constrained by renormalization condition to have a zero VEV.
}

{
Combining Figs.\ref{oneloop} and \ref{wave-re}, we {obtain} the $\lambda_3$-dependent
  one-loop contribution to the amplitude of VBF process.
  The {explicit expressions} will be given in section \ref{subsec:result}.
 }
% which are related with energy and observable, so the resulting modifications cannot be parameterised via a rescaling of the $VVh$ couplings in tree level~\cite{Degrassi:2016wml}.
% These contributions can ben classified into two categories: a process-dependent part proportional to $\lambda_{3}$, and a universal part proportional to $(\lambda_{3})^{2}$ from wave-function renormalisation of external Higgs boson.

%and \ref{wave-re}
%If we chose a different gauge, there would be an extra diagram with internal Goldstone lines.

\subsection{Top quark Yukawa coupling: $y_{t}$}
\begin{figure}[H]
\centering
\subfloat[]{\includegraphics[width=0.2\textwidth]{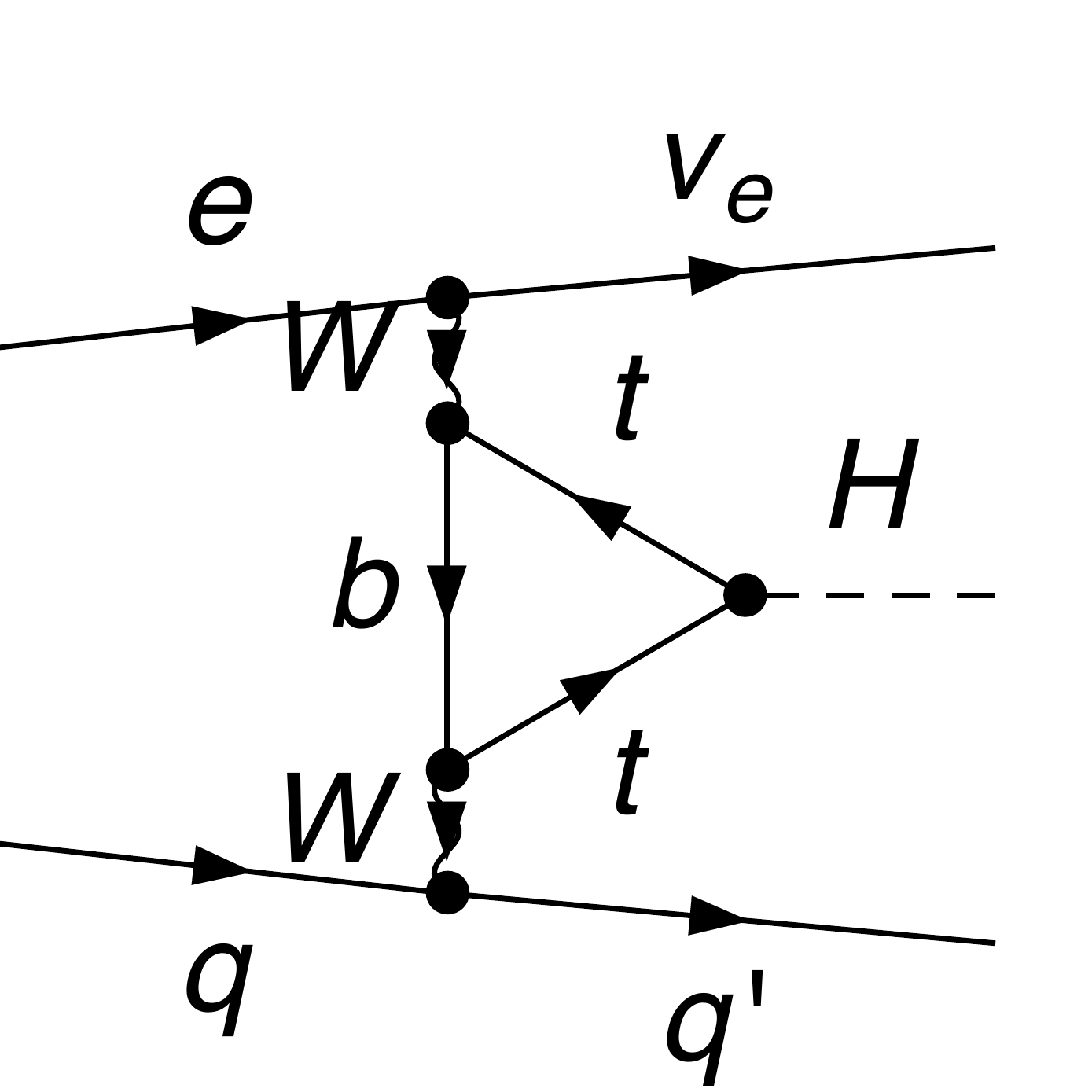}}
%\subfloat[]{\includegraphics[width=0.2\textwidth]{./top_triangle_2}}
\subfloat[]{\includegraphics[width=0.2\textwidth]{CT_HWW}}\\
\subfloat[]{\includegraphics[width=0.2\textwidth]{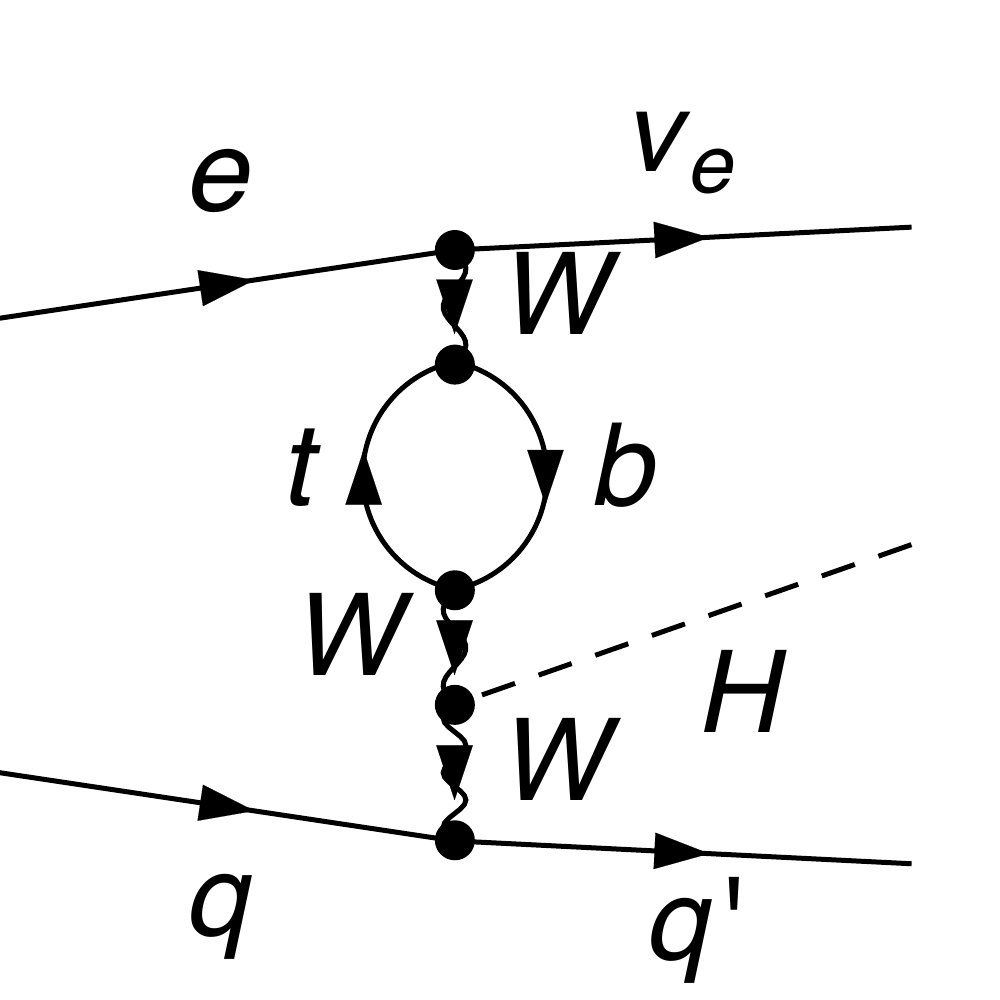}}
\subfloat[]{\includegraphics[width=0.2\textwidth]{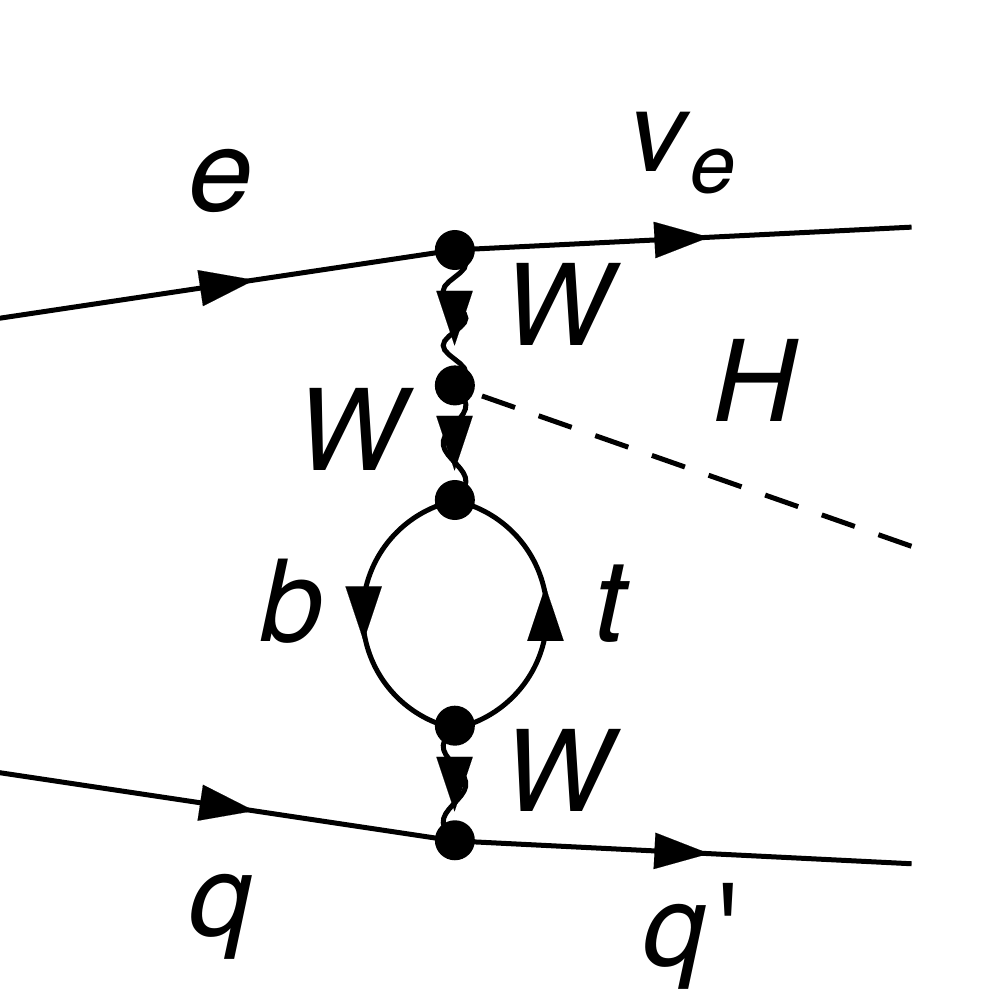}}
\subfloat[]{\includegraphics[width=0.2\textwidth]{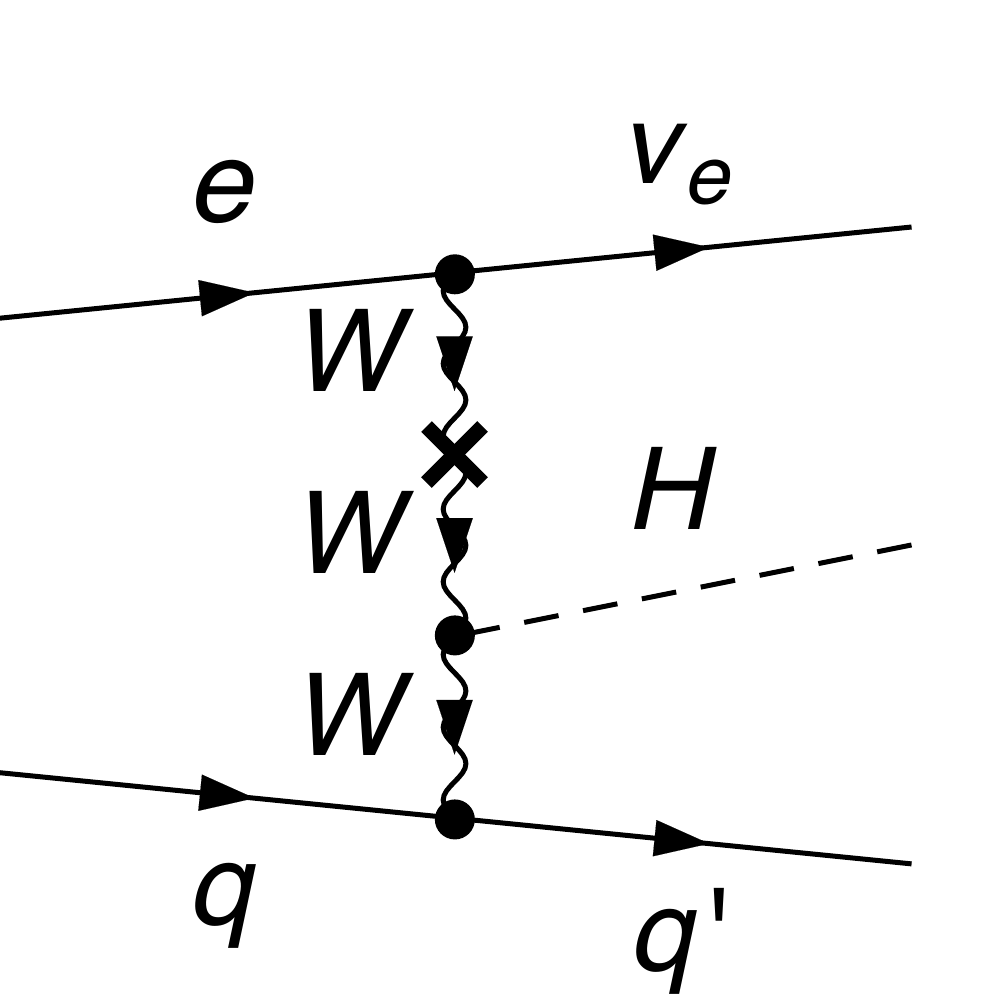}}
\subfloat[]{\includegraphics[width=0.2\textwidth]{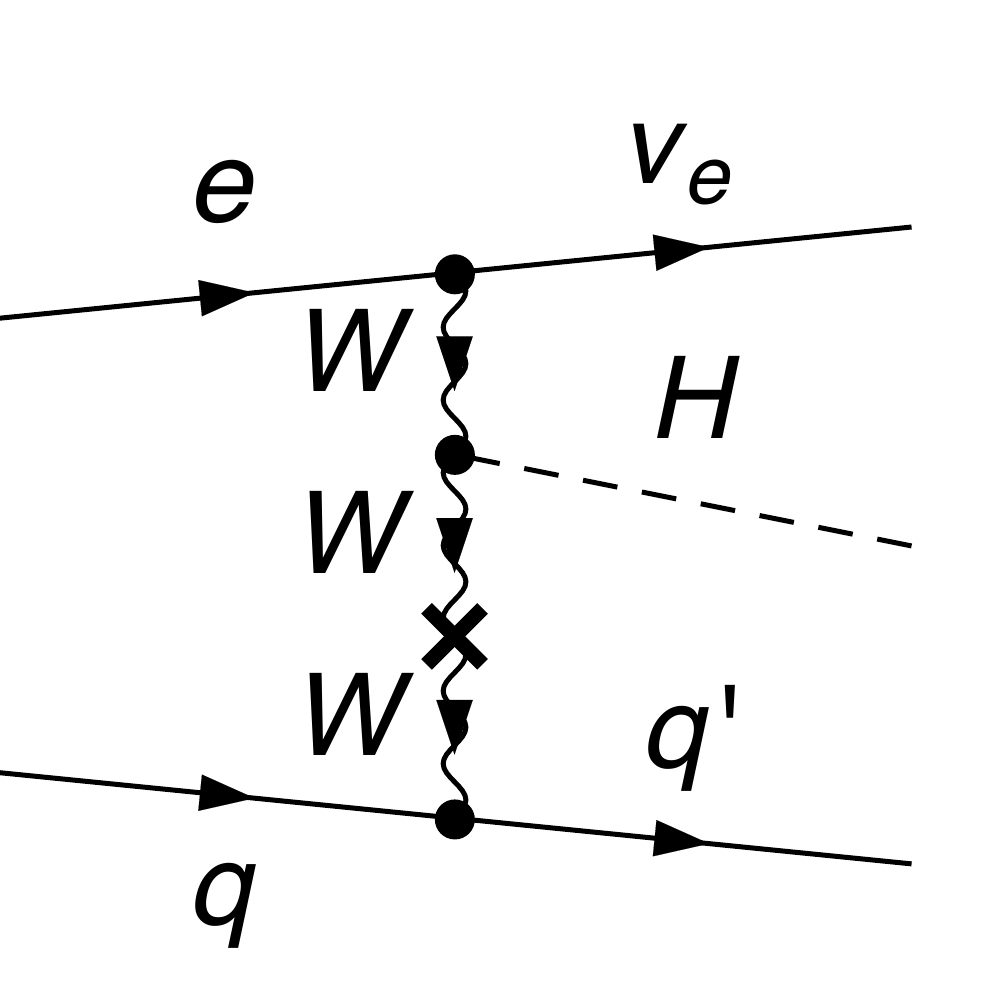}}\\
\subfloat[]{\includegraphics[width=0.2\textwidth]{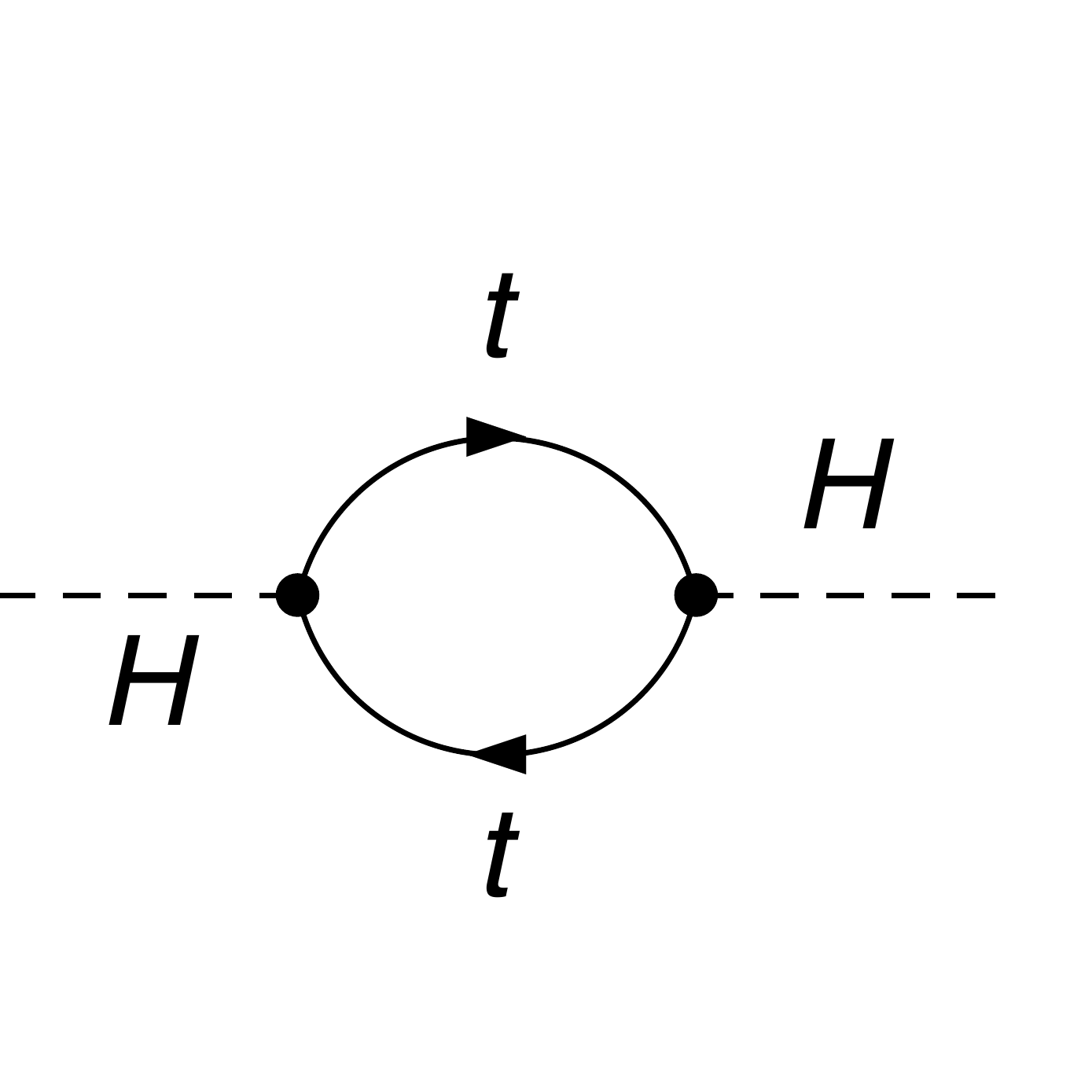}}
\subfloat[]{\includegraphics[width=0.2\textwidth]{wave_re_CT}}
%\subfloat[]{\includegraphics[width=0.25\textwidth]{./wavetop}}
	\caption{Feynman diagrams {with top and bottom loops} and the corresponding counter terms in the unitary gauge at one loop level with $q=u,c,\bar{d},\bar{s}$ and $q'=d,s,\bar{u},\bar{c}$.}
  %(left panel).
  %, and a universal part of wave-function renormalisation (right panel).}
\label{toploop}
\end{figure}

The same final states can be produced through the top quark Yukawa coupling to the Higgs boson.
 Because of the large coupling strength $y_{t}$ {and the insertion $M_{t}$ in the loop}, the contribution  from this channel
 may be sizable and could affect the determination of the trilinear Higgs self-coupling. Therefore, we shall treat it as an {irreducible background  and compute its contribution.} As it is not straightforward to separate the contribution from top and bottom quarks in the renormalization constants,  we calculate all the contribution of {top and bottom quarks} {by taking $\kappa_\lambda, e, M_H, M_W, M_Z $, {and $M_t$} as input parameters} in the $M_b \to 0$ limit. The Feynman diagrams in the unitary gauge are shown in Fig.\ref{toploop}. 
 (Diagrams that vanish in the $M_b \to 0$  limit are not shown here.) {The full analytic result will be shown in section \ref{subsec:result}.}

 %These graphs are independent of $\kappa_{\lambda}$.

%%%%%%%%%%%%%%%%%%%%%%%%%%%%%%%%%%%%%%%%%%%%
\subsection{Analytical result\label{subsec:result}}

In this section, we give the analytical result in both on-shell {and $\MSbar$ schemes. We shall use $\MSbar$ scheme in our numerical simulation in the next section.} 
{We do not show diagrams with Goldstone bosons in Figs.\ref{oneloop} and \ref{toploop}, but for the convenience of the calculation we restore those diagrams and work in Feynman gauge.}

We denote the {momenta} of the electron, incoming parton, Higgs boson, electron neutrino and outgoing parton {by} $p_{1-5}$ respectively.
The Mandelstam variables are defined as $S_{ij} \equiv (p_i + p_j)^2$, $T_{ij} \equiv (p_i - p_j)^2$.
In this work the {masses} of $u, c, d, s$ {quarks} are {neglected}.~The CKM matrix elements $V_{ub}, V_{cb}, V_{td}, V_{ts}$ are also taken {to be} zero.

We expand the amplitude $\mathcal{M}_{q}$ {of our process} {in powers of} 
$\gw\equiv \sqrt{\frac{4\pi\alpha_e}{\sin^2\theta_W}}$ as
\begin{equation}
  \label{eq:expand}
  \mathcal{M}_{q} = \gw^3\mathcal{M}_{q}^{(0)} + \gw^5\mathcal{M}_{q}^{(1)} + \cdots,
 %{~~\gw= \sqrt{\frac{4\pi\alpha_e}{\sin^2\theta_w}}},
\end{equation}
where $q$ stands for the incoming parton, 
$\alpha_e$ is the fine structure constant and $\theta_W \equiv \arccos{\frac{M_W}{M_Z}} $ is the Weinberg angle. The {squared amplitude is then} given by
\begin{align}
  \label{eq:ampsq}  
  & \frac{1}{2}\sum_{s} \frac{1}{3}\sum_{\text{color}}
    \left| \mathcal{M}_{q_s,e_L} \right|^2  \nn\\
  =&\frac{1}{8} \gw^6 M_W^2
    \frac{1}{( T_{14} - M_W^2)^2( T_{25} - M_W^2)^2}\FCA_q \nn\\
  +& \frac{3 M_H^2}{128 \pi^2} \gw^8
     \frac{1}{( T_{14} - M_W^2)^2( T_{25} - M_W^2)^2} \times G_\lambda \nn\\
%%%%%%%%%%%%%%%%%%%%%%%%%%%%%%%%%%%%%%%%%%%%%%%%%%%%%%%
% t+b: triangle diag 
%%%%%%%%%%%%%%%%%%%%%%%%%%%%%%%%%%%%%%%%%%%%%%%%%%%%%%%
  - &\frac{N_c M_t^2}{128 \pi^2}\gw^8
     \frac{1}{(T_{14} - M_W^2)^2( T_{25} - M_W^2)^2} \times G^{(3)}_{t+b}\nn\\
%%%%%%%%%%%%%%%%%%%%%%%%%%%%%%%%%%%%%%%%%%%%%%%%%%%%%%%
% t+b: W prop. self-energy
%%%%%%%%%%%%%%%%%%%%%%%%%%%%%%%%%%%%%%%%%%%%%%%%%%%%%%%
  + &\left[
  \frac{N_c M_W^2}{128 \pi^2}\gw^8
     \frac{1}{(T_{14} - M_W^2)^3( T_{25} - M_W^2)^2} \times G^{(14)}_{t+b}
  + 14 \leftrightarrow 25
  \right] \\
  +& \cdots, \nn
\end{align}
{where
  $s$ is spin of the incoming quark(anti-quark),
  the subscripts of $\mathcal{M}$ denote the flavor and
  spin of the incoming quark(anti-quark) and electron.
  The polarisation of the electron beam will be considered in next section.
}
The fermion chains $\mathcal{F}_q^{(1)}$ are
\begin{align}
  \label{eq:FC}
  \FCA_{u,c} &= 4 S_{12} S_{45} \nn\\
  \FCA_{\bar{d},\bar{s},\bar{b}} &= 4  T_{15} T_{24} \nn\\
  \FCB_{u, c} &= (S_{12} + S_{45})(S_{12} S_{45} + T_{15} T_{24} - T_{14} T_{25}) 
  + 2 S_{12} S_{45} (T_{15} + T_{24}) \nn\\
  \FCB_{\bar{d},\bar{s},\bar{b}} &=  (T_{15} + T_{24})(S_{12} S_{45} + T_{15} T_{24} - T_{14} T_{25})
  + 2 T_{15} T_{24} (S_{12} + S_{45}) ,
\end{align}

where $q=u,c,\bar{d}, \bar{s}, \bar{b}$ is the incoming parton. The $\kappa_{\lambda}$ dependent term $G_\lambda$ reads {
\begin{align}
  \label{eq:GH}
G_\lambda = 
%    \left\{
	&\left[
    \left(C_{00} - \frac{1}{4}B_0 - M_W^2 C_0\right) \FCA_q
	- C_{12} \FCB_q 
    \right] \kappa_{\lambda} \nn \\
    &-\frac{3}{8}\kappa_{\lambda}^2 M_H^2 B'_0 \FCA_q.
%    \right\}
\end{align}
}
{Note that the contribution from $\kappa_{\lambda}$ dependent counter terms vanishes. $G_\lambda$ takes the same form in both OS and $\MSbar$ schemes, but the quantities on which it depends are generally renormalization scheme dependent.}
{The} contribution from top {and bottom quarks}, in the $M_b \to 0$ limit, is given by
\begin{align}
  G^{(3)}_{t+b}
  &=
%%%%%%%%%%%%%%%%%%%%%%%%%%%%%%%%%%%%%%%%%%%%%%%%%%%%%%%  
%Fermion chain 1
%%%%%%%%%%%%%%%%%%%%%%%%%%%%%%%%%%%%%%%%%%%%%%%%%%%%%%%  
% the loop diag
    \kappat \Big\{
     4 \ct_{00} - 2 B_0(T_{25},M_b^2,M_t^2)
    - 2(M_t^2 - T_{14}) \ct_0 
    + \half\left( M_H^2 + 5 T_{14} - T_{25} \right) \ct_1 \nn\\
  &~~~~  + \half \left( - 3 M_H^2 + 3 T_{14} + T_{25} \right) \ct_2 \nn\\
% the wave-function corr.  
  &~~~~ - \half \left[ (4 M_t^2 - M_H^2)\btp_{0}  -\bt_{0} \right] \kappat \nn\\
% the counter-term corr. of both vect. and wave-func. corr.    
  &~~~~  + \text{CT}^{(3)}_{t+b}
    \Big\} \FCA_q     \nn \\
%%%%%%%%%%%%%%%%%%%%%%%%%%%%%%%%%%%%%%%%%%%%%%%%%%%%%%%  
%Fermion chain 2
%%%%%%%%%%%%%%%%%%%%%%%%%%%%%%%%%%%%%%%%%%%%%%%%%%%%%%%  
  &~~ - \kappat (\ct_1 + \ct_2 + 4 \ct_{12}) \FCB_q\nn\\
%%%%%%%%%%%%%%%%%%%%%%%%%%%%%%%%%%%%%%%%%%%%%%%%%%%%%%%  
% W prop. self energy
%%%%%%%%%%%%%%%%%%%%%%%%%%%%%%%%%%%%%%%%%%%%%%%%%%%%%%%
  G^{(14)}_{t+b}
  &=\left[ -4B_{00}(T_{14},M_b^2,M_t^2)
    + 2(M_t^2 - T_{14})B_0(T_{14},M_b^2, M_t^2)
    - 2T_{14}B_1(T_{14},M_b^2,M_t^2) + \text{CT}^{(14)}_{t+b}
    \right] \FCA_q 
%    \biggr\} \nn\\
\end{align}
with contributions from counter terms \cite{Denner:1991kt}
\begin{align}
  % see MMA
  %%%%%%%%%%%%%%%%%%%%%%%%%%%%%%%%%%%%%%%%%%%%%%%%%%%%%%% 
  % OS scheme
  %%%%%%%%%%%%%%%%%%%%%%%%%%%%%%%%%%%%%%%%%%%%%%%%%%%%%%% 
  \text{CT}^{(3)}_{t+b}|_\text{OS}
  % 2 LO * NLO^*
  =&2 \frac{-16\pi^2}{N_c \gw^2}
  \frac{M_W^2}{M_t^2}
  \left[
  %the HWW CT
  \left(
  \delta Z_e - \frac{\delta s}{s} + \frac{1}{2} \frac{\delta M_W^2}{M_W^2}
  + \frac{1}{2}\delta Z_H + \delta Z_W \right)
  %the wave-func corr. CT
	{{- \frac{1}{2}\delta Z_H}}  
  \right]_{t+b} \nn\\
  %%%%%%%%%%%%%%%%%%%%%%%%%%%%%%%%%%%%%%%%%%%%%%%%%%%%%%% 
  % MSbar 
  %%%%%%%%%%%%%%%%%%%%%%%%%%%%%%%%%%%%%%%%%%%%%%%%%%%%%%% 
  \text{CT}^{(3)}_{t+b}\big|_{\MSbar}
	=&\frac{1}{2}\left( \frac{2}{4-D}\right) \nn\\
  %%%%%%%%%%%%%%%%%%%%%%%%%%%%%%%%%%%%%%%%%%%%%%%%%%%%%%% 
  % CT14: OS
  %%%%%%%%%%%%%%%%%%%%%%%%%%%%%%%%%%%%%%%%%%%%%%%%%%%%%%% 
  \text{CT}^{(14)}_{t+b}|_\text{OS}
  =& \frac{32 \pi^2}{N_C \gw^2} (-1)
	 \left[ \delta Z_W (T_{14} - M_W^2) - \delta M_W^2 \right]_{t+b} \nn\\
  %%%%%%%%%%%%%%%%%%%%%%%%%%%%%%%%%%%%%%%%%%%%%%%%%%%%%%% 
  % CT14: MSbar 
  %%%%%%%%%%%%%%%%%%%%%%%%%%%%%%%%%%%%%%%%%%%%%%%%%%%%%%% 
	 \text{CT}^{(14)}_{t+b}|_{\MSbar}
	=& \left(\frac{2}{3}T_{14}- M_t^2 \right)\left( \frac{2}{4-D}\right)
\end{align}
where the subscript $t+b$ represents the contribution from the top and bottom quarks, $D$ is the dimension of space-time and $\delta Z_e, \delta Z_H, \delta Z_W, \delta s, \delta M_W^2$ are renormalization constants in {the} on-shell scheme. {Detailed} expressions of these renormalization constants can be found in \cite{Denner:1991kt} 
(Note that, at one loop level, all the renormalization constants in the above equations are independent of $\kappa_\lambda$
except for $\delta Z_H$, whose contribution cancels out in the final result.  
%The contribution of $\delta Z_H$ cancels out in the final result, 
%we do not bother to write down its detailed expression here.
)
%\sout{(It can be shown that variation of the Higgs self-coupling does not change the form of these renormalization constants).}  
$B_0, B'_0, C_x$(e.g. $C_{00},C_{1}$, \text{etc.}) are scalar integrals
\begin{align}
  \label{eq:PVhiggs}
  C_{x} &= C_{x}(T_{14}, M_H^2, T_{25}, M_W^2, M_H^2, M_H^2) \nn \\
  B_0 &= B_0(M_H^2, M_H^2,M_H^2) \nn\\
  B'_0 &= \frac{\partial}{\partial s}B_0(s, M_H^2,M_H^2)\Big|_{s = M_{H,\text{ph}}^2}
% valid for OS scheme only:
%   = \left( \frac{2\pi}{3\sqrt{3}} - 1 \right) \frac{1}{M_H^2}
\end{align}
and
\begin{align}
  \label{eq:b0t}
  \ct_{x} &= C_{x}(T_{14}, M_H^2, T_{25},M_b^2, M_t^2, M_t^2) \nn\\
  \bt_{0} &= B_0(M_H^2, M_t^2, M_t^2)    \nn\\
  \btp_{0} &= \frac{\partial}{\partial s}B_0(s, M_t^2,M_t^2)\Big|_{s = M_{H,\text{ph}}^2}.
%  B_0^{(2)} &= B_0(T_{25}, M_b^2, M_t^2)
\end{align}
Here $M_{H,\text{ph}}$ is the physical mass of Higgs boson
while values of other mass parameters, e.g. $M_H$, depend on the renormalization scheme used.

%The {\it FeynArts}, {\it FormCalc} and {\it LoopTools} suite of packages~\cite{Hahn:2000kx,Hahn:1998yk} is used to {produce the analytical result in this section and the numerical result in Sec.~\ref{MC}}. 

{As a cross check, we repeat the calculation with {\it FeynCalc}~\cite{Mertig:1990an,Shtabovenko:2016sxi} and obtain the same result.~Gauge invariance is verified by working in the general $R_{\xi}$ gauge and making sure that no gauge parameter dependence remains in the final result.} 
%%%%%%%%%%%%%%%%%%%%%%%%%%%%%%%%%%%%%%%%%%%%

\section{\label{MC}Monte Carlo simulation}
\begin{figure}[H]
\centering
{\includegraphics[scale=0.42]{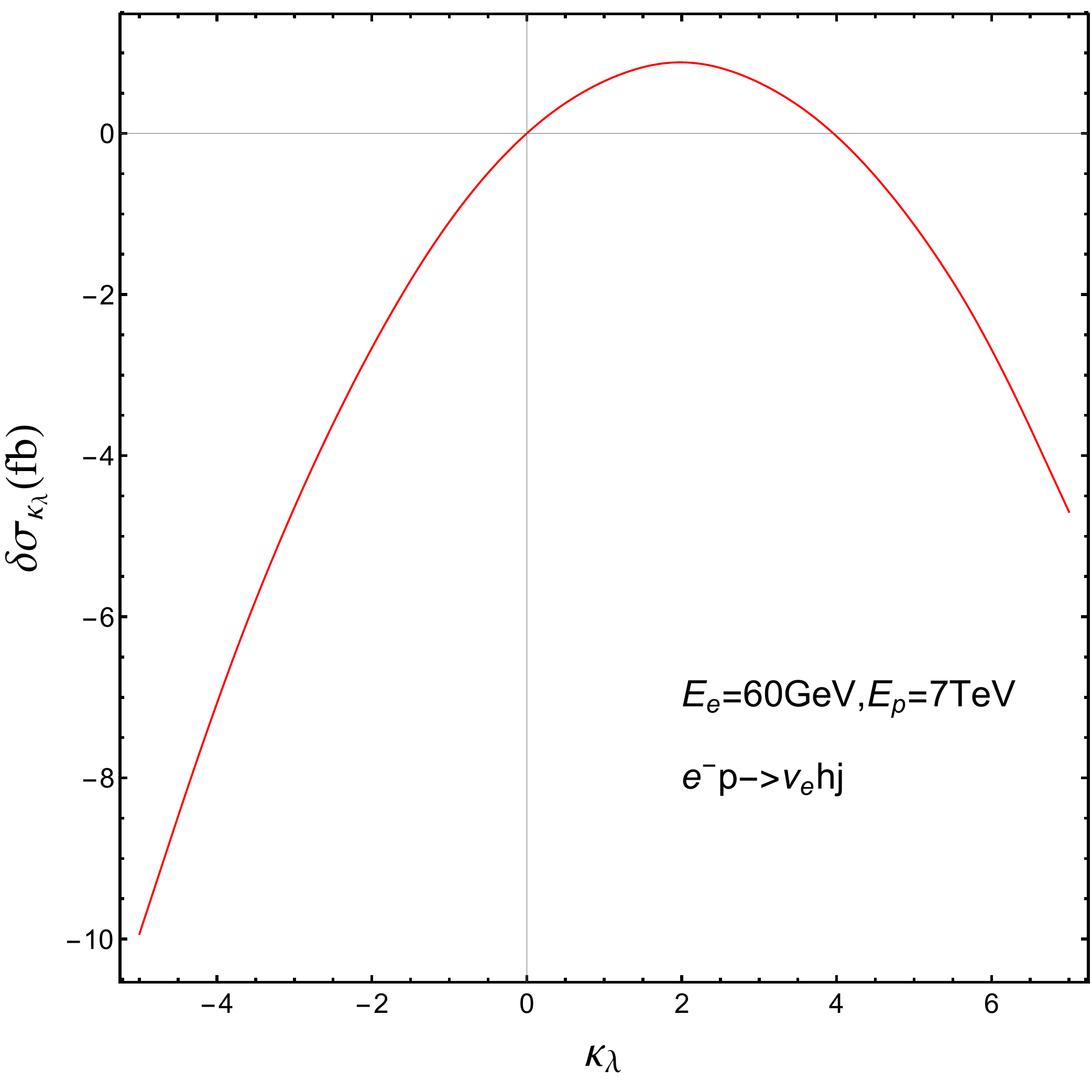}}
\caption{The cross section $\delta\sigma_{\kappa_\lambda}$ from the $\kappa_{\lambda}$ dependent one-loop corrections varying with  $\kappa_{\lambda}$,
  {with basic cuts given in Eq.\ref{eq:cuts}. The electron polarisation is taken to be -80\%}.}
\label{higgsinterfere}
\end{figure}

{The squared amplitude in Eq.\ref{eq:ampsq} can be turned into {the} NLO cross section $\sigma_{\kappa_{\lambda}}$ for the process $e^{-}p \to \nu_{e} h j$ {after integration over the phase space of the final states {and convolution with parton distribution function (PDF) of the incoming quark(anti-quark)}.}
~We use {\it Vegas} algorithm implemented in the Cuba library~\cite{Hahn:2004fe} to {perform the numerical integration in our simulation}} at parton-level.
{~We also promote it to a parton level event generator to study the variation of kinematical variables.}
~{Given the numerical values of the input parameters, the scalar integrals $B_0, B'_0, C_x$(e.g. $C_{00},C_{1}$, \text{etc.}) are evaluated with the {\it LoopTools}~\cite{Hahn:1998yk} package.}
{The PDF set {\it CT14qed\_inc\_proton} is used~\cite{Schmidt:2015zda}. Both the renormalization and factorization scales are set to $M_H=125~\text{GeV}$.}~The {following basic cuts are adopted:}
\bea
\label{eq:cuts}
&&p_{T}^{j}>20~\text{GeV}, p_{T}^{\ell}>5~\text{GeV} \nonumber \\
&&|\eta_{j}|<5,|\eta_{\ell}|<5\\
&&\slashed{E}_{T}>10~\text{GeV}. \nonumber 
\eea
{With beam energies being $7~\text{TeV}$ and $60~\text{GeV}$ for the proton and electron,}
{and eletron polarisation assumed to be -80\%},
the cross section of the {process} $e^{-}p \to \nu_{e} h j$ is {145}~fb at leading order.
In Fig.\ref{higgsinterfere}, we show the cross section $\delta\sigma_{\kappa_\lambda}$ from only the $\kappa_{\lambda}$ dependent one loop terms.
{The quadratic form can be traced back to the $\kappa_{\lambda}$ and $(\kappa_{\lambda})^{2}$ terms in Eq.\ref{eq:GH}.}
The universal contribution from the top and bottom quarks turns out to be {-1.8}~fb.
The {\it Vegas} phase space integration is cross checked with {\it MadGraph5\_v2.6.5}~\cite{Alwall:2014hca}, which gives the consistent result.

\begin{figure}[H]
\centering
\subfloat[The azimuthal angle $\Delta\phi_{\slashed{E}_{T}j}$]{\includegraphics[scale=0.37]{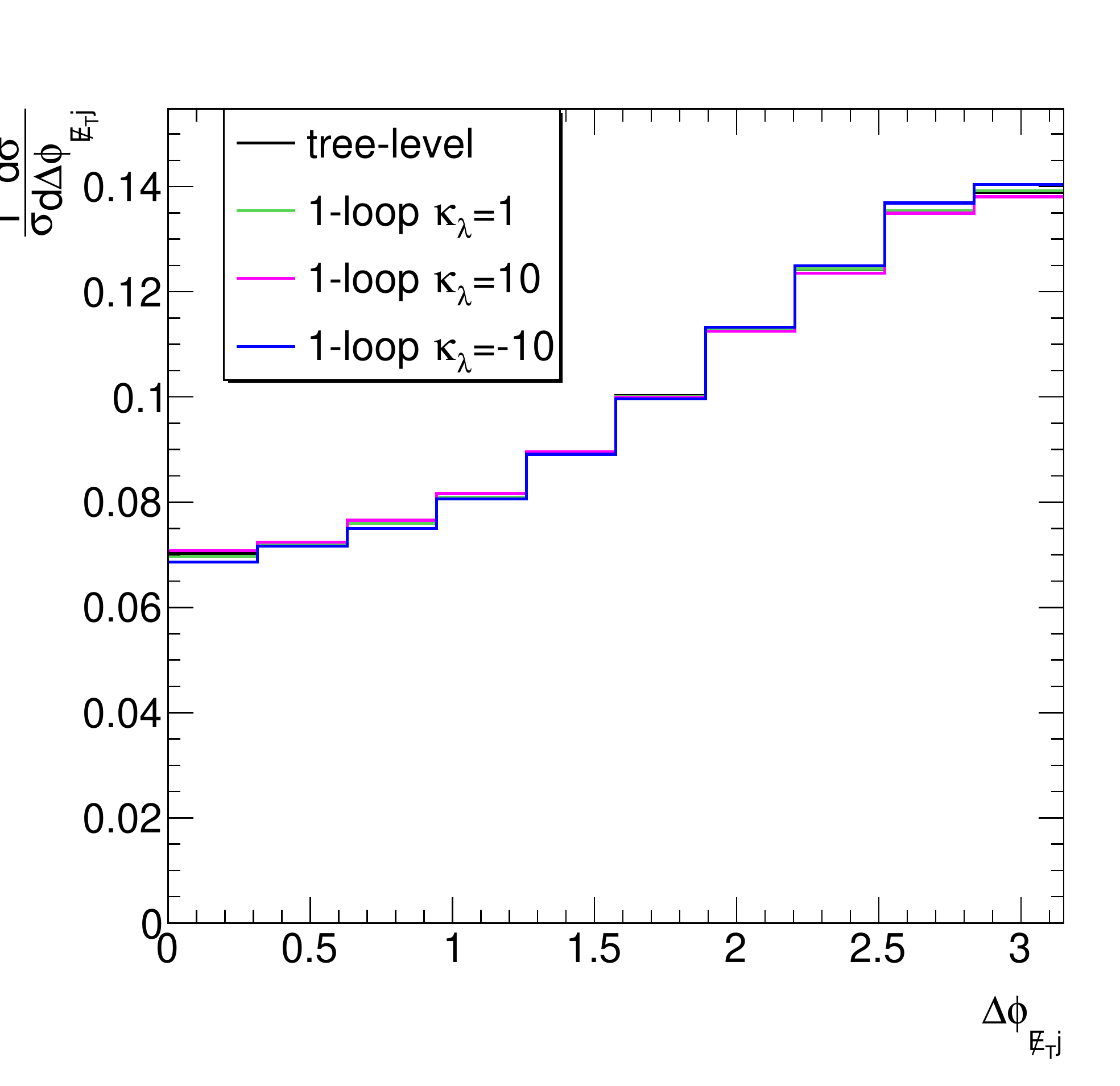}} \qquad
\subfloat[The Higgs transverse momentum $p_{T}^{h}$]{\includegraphics[scale=0.37]{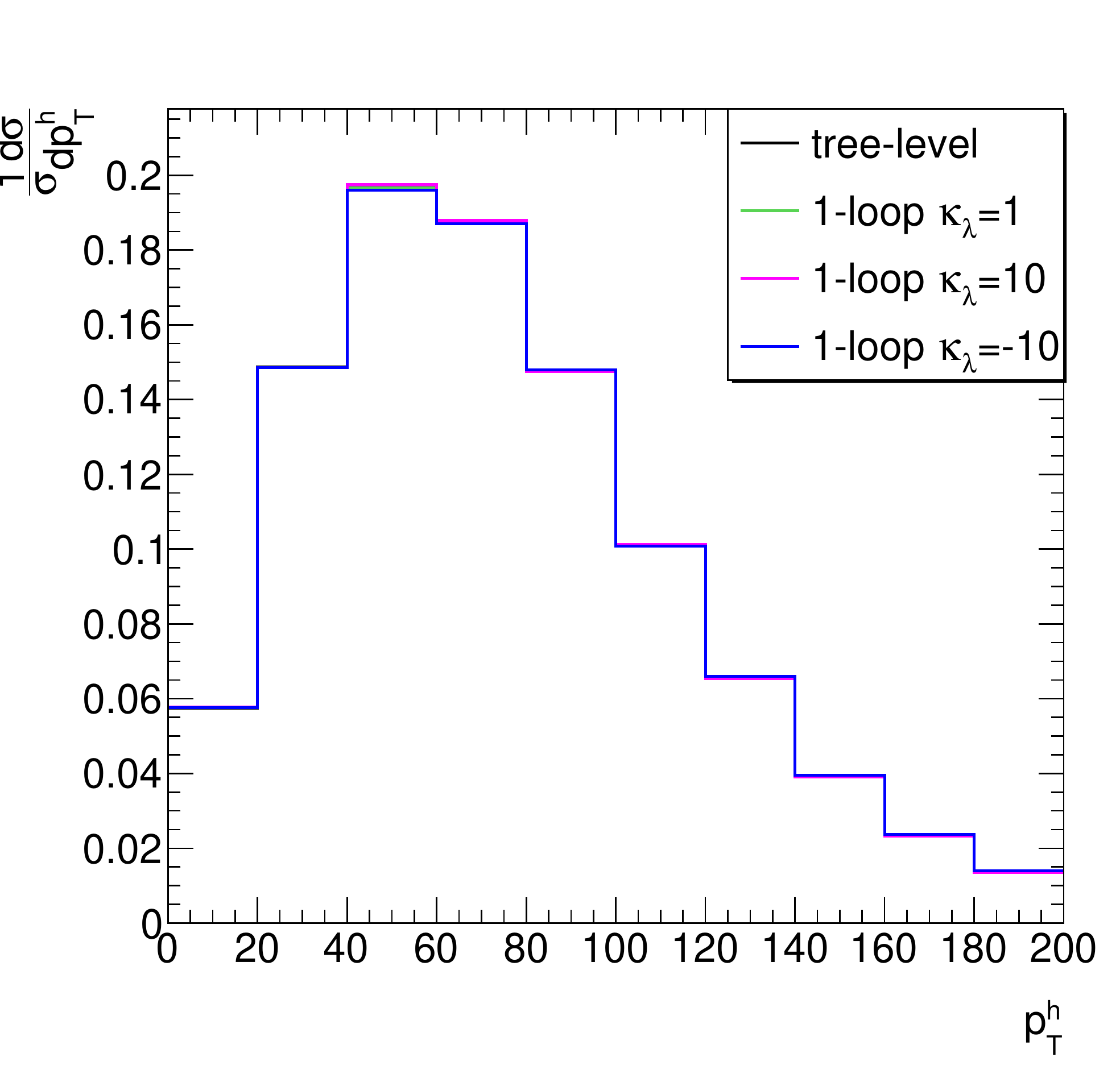}}
	\caption{The normalized $\Delta\phi_{\slashed{E}_{T}j}$ (left panel) and $p_{T}^{h}$ (right panel) distributions {with various} $\kappa_{\lambda}$.}
\label{distribution}
\end{figure}

{One way to show the significance of $\kappa_{\lambda}$ is via the differential distributions of characteristic kinematic variables, such as  the azimuthal angle $\Delta\phi_{\slashed{E}_{T}j}$, the Higgs transverse momentum $p_{T}^{h}$, etc.} {Unfortunately, the discrimination between distributions for various processes relies heavily on the effect of threshold Sommerfeld enhancement, which is absent in the case of loop corrections with the Higgs trilinear self-coupling~\cite{Maltoni:2017ims}. This is very well illustrated in Fig.\ref{distribution} even when $\kappa_{\lambda}$ is varied in a very wide range. The distributions are normalized to reflect only the difference in shape.}

{As there is little difference in shapes of distributions, we seek to identify the anomalous Higgs self interaction from SM processes using their {production rate}.~This can be done with the $\chi^{2}$ method, in which the deviation between BSM and SM cross sections are described by}
% \bea
% \chi^{2}=(\frac{\sigma_{\kappa_{\lambda}\neq1}-\sigma_{\kappa_{\lambda}=1}}{\sqrt{\sigma_{\kappa_{\lambda}=1}}})^{2} \cdot \mathcal{L},
% \label{chi2method}
% \eea
\begin{equation}
  \chi^{2}=\left( \frac{N_{\kappa_{\lambda}\neq 1}-N_{\kappa_{\lambda}=1}}{\sqrt{N_{\kappa_{\lambda}=1}}} \right)^{2},
 \label{chi2method}
\end{equation}
where $N_{\kappa_{\lambda}=1}$ is the number of events of the $e^{-}p \to \nu_{e} h j$ with $\kappa_{\lambda}=1$,
while $N_{\kappa_{\lambda}\neq 1}$ contains the anomalous $\kappa_{\lambda}$ contribution. 
The number of events is given by
 \begin{equation}
   N_{\kappa_\lambda} = \mathcal{L} \cdot \sigma_{\kappa_\lambda} \cdot \text{BR}_{\kappa_\lambda}
   %\cdot \epsilon + \text{background}
 \end{equation}
where $\mathcal{L}$ is the integrated luminosity, BR is the branch ratio.
%  $\epsilon$ is the cut efficiency.
The universal contribution from the top and bottom quarks is also included.
The dependence of branch ratio on $\kappa_\lambda$ has been studied in ~\cite{Degrassi:2016wml}.
% \blue{Based on the analysis of differential distributions,
%   $\epsilon$ is assumed to be indepnedent of $\kappa_\lambda$}
We apply the $\chi^{2}$ method to the $H \to b\bar{b}$ decay channel for its large branch ratio ($BR(h \to b\bar{b})\approx58\%$)~\cite{Tanabashi:2018oca}.
The {limits on $\kappa_{\lambda}$ at 95\% C.L. are } listed in Table.\ref{constraints}.
We find {that $\kappa_{\lambda}$ is better constrained} with the increase of the integrated luminosity.
The most stringent limits on $\kappa_{\lambda}$ is {[-0.42, 2.79]},
{with} $\mathcal{L}=3~\text{ab}^{-1}$.

\begin{table}[H]
\renewcommand\arraystretch{0.7}
\centering
\begin{tabular}{|c||c|}
\hline
\hline
The integrated luminosity & Bounds of the $\kappa_{\lambda}$  \\
\hline
\hline
$\mathcal{L}=1~\text{ab}^{-1}$ & [-0.88, 3.33]  \\
\hline
$\mathcal{L}=2~\text{ab}^{-1}$ & [-0.57, 2.98]  \\
\hline
$\mathcal{L}=3~\text{ab}^{-1}$ & [-0.42, 2.79]  \\
\hline
\hline
\end{tabular}
	\caption{{The 95\% C.L. bounds on $\kappa_{\lambda}$  for various integrated luminosities.} }
\label{constraints}
\end{table} 

The above results are obtained {at parton-level} without considering any background. Fortunately, deep inelastic scattering machines, e.g. LHeC, and FCC-eh have great potential for high precision Higgs physics. {The measurement of $H \to b\bar{b}$ has reached a precision of $\mathcal{O}(1\%)$ \cite{Klein:2018rhq,Abada:2019lih},} which is good enough for probing the Higgs self-coupling at NLO.
%If we use the result from the cut-based analysis in \cite{Han:2009pe,Mashiro:2017aqd}
{Refs.}~{\cite{Han:2009pe,Mashiro:2017aqd} has studied the single Higgs production and its background {in} detail. In \cite{Mashiro:2017aqd}, both cut based and BDT based analyses are pursued, and reasonable detector setups for LHeC are considered. Combining the {efficiencies of the cut and Heavy flavor tagging,} the survival rate of signal is about $7\%\sim10$\%, and {the ratio of the surviving signal and background is as large as 2.9.} If we use these efficiencies to estimate the event numbers at the detector level,} the {bounds in Table.\ref{constraints}} are then broadened to the {values} shown in Table.\ref{constraintsafterdet}. {(The 1-loop QCD correction to VBF processes at percent level~\cite{Jager:2010zm} is not included, which has negligible effect on our result ).}

\begin{figure}[H]
\centering
{\includegraphics[scale=0.42]{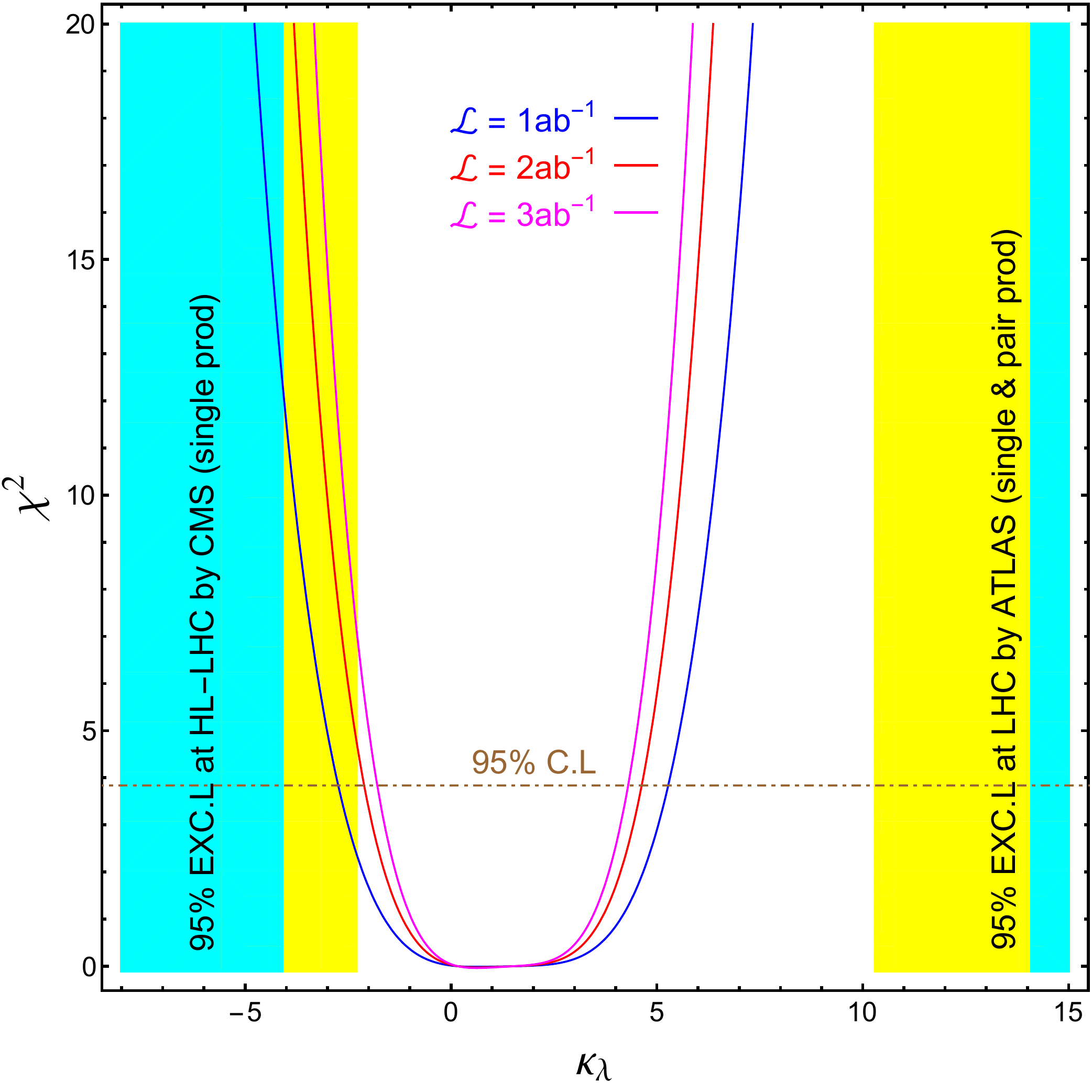}}
	\caption{{The 95\% C.L. bounds on $\kappa_{\lambda}$ {for various} integrated luminosities and current experimental constraints. }}
\label{bound}
\end{figure}

The bounds are {also} shown in Fig.\ref{bound}, in which the solid blue, red and magnet curves correspond to different integrated luminosities, the cyan shade represents the projection of 95\% C.L. exclusion at the HL-LHC with 3 ab$^{-1}$ by CMS via single Higgs production \cite{Cepeda:2019klc}, and the yellow shade represents the recent 95\% C.L. exclusion from combination of double and single Higgs production by ATLAS~\cite{ATLAS:2019pbo}.
 ~{The signal exclusion at the LHeC is clearly more efficient.}
 {However, it should be noted that the analysis of double Higgs production at the HL-LHC gives a better prospective constraint [0.1, 2.3]~\cite{Cepeda:2019klc}}.
 Our results can be improved if the background in the measurement could be further reduced with more sophisticated methods (e.g., the machine learning, {which is {helpful in} heavy flavor tagging \cite{Abada:2019lih}}).

\begin{table}[H]%\footnotesize
\renewcommand\arraystretch{0.7}
\centering
\begin{tabular}{|c||c|}
\hline
\hline
The integrated lnuminosity & Bounds of the $\kappa_{\lambda}$  \\
\hline
\hline
$\mathcal{L}=1~\text{ab}^{-1}$ & [-2.74, 5.28]  \\
\hline
$\mathcal{L}=2~\text{ab}^{-1}$ & [-2.11, 4.63]  \\
\hline
$\mathcal{L}=3~\text{ab}^{-1}$ & [-1.79, 4.30]  \\
\hline
\hline
\end{tabular}
	\caption{{The 95\% C.L. bounds on $\kappa_{\lambda}$  for various integrated luminosities after the cut-based analysis.} }
\label{constraintsafterdet}
\end{table} 

\section{\label{con}Conclusion}
In this paper, we study the significance of the Higgs trilinear coupling through the single Higgs production process $e^{-}p \to \nu_{e} h j$ at the LHeC.
  The analytical calculation is carried out up to one loop level for both $\lambda_3$ dependent and independent intermediate states.
  The analytical results are then used in the Monte Carlo simulation to produce numerically the cross section at various $\kappa_{\lambda}$,
  which allows us to quantify the deviation of the cross section from the SM case ($\kappa_{\lambda}=1$) in a $\chi^{2}$ statistic analysis.
  From this analysis we find that the 95\% C.L. bound for $\kappa_{\lambda}$ is {[-2.11, 4.63]} with a 2 ab$^{-1}$ integrated luminosity,
  {after using the result from an existing cut-based analysis.}
  ~This is a significant improvement compared with the current experimental result.
  We expect the result to be improved with more accurate measurement of Higgs decays.

\section*{Acknowledgments}
We thank Dr. Chen Shen, Prof. Uta Klein and LHeC Higgs \&Top group for helpful discussion. The work is supported in part by the National Science Foundation of China~(11875232)~and the Zhejiang University Fundamental Research Funds for the Central Universities. KW is also supported by Zhejiang University K.P Chao High Technology Development Foundation.

\bibliographystyle{apsrev4-1}
%\bibliography{THC}

\begin{thebibliography}{74}%
\makeatletter
\providecommand \@ifxundefined [1]{%
 \@ifx{#1\undefined}
}%
\providecommand \@ifnum [1]{%
 \ifnum #1\expandafter \@firstoftwo
 \else \expandafter \@secondoftwo
 \fi
}%
\providecommand \@ifx [1]{%
 \ifx #1\expandafter \@firstoftwo
 \else \expandafter \@secondoftwo
 \fi
}%
\providecommand \natexlab [1]{#1}%
\providecommand \enquote  [1]{``#1''}%
\providecommand \bibnamefont  [1]{#1}%
\providecommand \bibfnamefont [1]{#1}%
\providecommand \citenamefont [1]{#1}%
\providecommand \href@noop [0]{\@secondoftwo}%
\providecommand \href [0]{\begingroup \@sanitize@url \@href}%
\providecommand \@href[1]{\@@startlink{#1}\@@href}%
\providecommand \@@href[1]{\endgroup#1\@@endlink}%
\providecommand \@sanitize@url [0]{\catcode `\\12\catcode `\$12\catcode
  `\&12\catcode `\#12\catcode `\^12\catcode `\_12\catcode `\%12\relax}%
\providecommand \@@startlink[1]{}%
\providecommand \@@endlink[0]{}%
\providecommand \url  [0]{\begingroup\@sanitize@url \@url }%
\providecommand \@url [1]{\endgroup\@href {#1}{\urlprefix }}%
\providecommand \urlprefix  [0]{URL }%
\providecommand \Eprint [0]{\href }%
\providecommand \doibase [0]{http://dx.doi.org/}%
\providecommand \selectlanguage [0]{\@gobble}%
\providecommand \bibinfo  [0]{\@secondoftwo}%
\providecommand \bibfield  [0]{\@secondoftwo}%
\providecommand \translation [1]{[#1]}%
\providecommand \BibitemOpen [0]{}%
\providecommand \bibitemStop [0]{}%
\providecommand \bibitemNoStop [0]{.\EOS\space}%
\providecommand \EOS [0]{\spacefactor3000\relax}%
\providecommand \BibitemShut  [1]{\csname bibitem#1\endcsname}%
\let\auto@bib@innerbib\@empty
%</preamble>
\bibitem [{\citenamefont {Aad}\ \emph {et~al.}(2012)\citenamefont {Aad} \emph
  {et~al.}}]{Aad:2012tfa}%
  \BibitemOpen
  \bibfield  {author} {\bibinfo {author} {\bibfnamefont {G.}~\bibnamefont
  {Aad}} \emph {et~al.} (\bibinfo {collaboration} {ATLAS}),\ }\href {\doibase
  10.1016/j.physletb.2012.08.020} {\bibfield  {journal} {\bibinfo  {journal}
  {Phys. Lett.}\ }\textbf {\bibinfo {volume} {B716}},\ \bibinfo {pages} {1}
  (\bibinfo {year} {2012})},\ \Eprint {http://arxiv.org/abs/1207.7214}
  {arXiv:1207.7214 [hep-ex]} \BibitemShut {NoStop}%
%%CITATION = ARXIV:1207.7214;%%
\bibitem [{\citenamefont {Chatrchyan}\ \emph {et~al.}(2012)\citenamefont
  {Chatrchyan} \emph {et~al.}}]{Chatrchyan:2012xdj}%
  \BibitemOpen
  \bibfield  {author} {\bibinfo {author} {\bibfnamefont {S.}~\bibnamefont
  {Chatrchyan}} \emph {et~al.} (\bibinfo {collaboration} {CMS}),\ }\href
  {\doibase 10.1016/j.physletb.2012.08.021} {\bibfield  {journal} {\bibinfo
  {journal} {Phys. Lett.}\ }\textbf {\bibinfo {volume} {B716}},\ \bibinfo
  {pages} {30} (\bibinfo {year} {2012})},\ \Eprint
  {http://arxiv.org/abs/1207.7235} {arXiv:1207.7235 [hep-ex]} \BibitemShut
  {NoStop}%
%%CITATION = ARXIV:1207.7235;%%
\bibitem [{\citenamefont {Aad}\ \emph {et~al.}(2015{\natexlab{a}})\citenamefont
  {Aad} \emph {et~al.}}]{Aad:2015zhl}%
  \BibitemOpen
  \bibfield  {author} {\bibinfo {author} {\bibfnamefont {G.}~\bibnamefont
  {Aad}} \emph {et~al.} (\bibinfo {collaboration} {ATLAS, CMS}),\ }\href
  {\doibase 10.1103/PhysRevLett.114.191803} {\bibfield  {journal} {\bibinfo
  {journal} {Phys. Rev. Lett.}\ }\textbf {\bibinfo {volume} {114}},\ \bibinfo
  {pages} {191803} (\bibinfo {year} {2015}{\natexlab{a}})},\ \Eprint
  {http://arxiv.org/abs/1503.07589} {arXiv:1503.07589 [hep-ex]} \BibitemShut
  {NoStop}%
%%CITATION = ARXIV:1503.07589;%%
\bibitem [{\citenamefont {Khachatryan}\ \emph {et~al.}(2015)\citenamefont
  {Khachatryan} \emph {et~al.}}]{Khachatryan:2014kca}%
  \BibitemOpen
  \bibfield  {author} {\bibinfo {author} {\bibfnamefont {V.}~\bibnamefont
  {Khachatryan}} \emph {et~al.} (\bibinfo {collaboration} {CMS}),\ }\href
  {\doibase 10.1103/PhysRevD.92.012004} {\bibfield  {journal} {\bibinfo
  {journal} {Phys. Rev.}\ }\textbf {\bibinfo {volume} {D92}},\ \bibinfo {pages}
  {012004} (\bibinfo {year} {2015})},\ \Eprint {http://arxiv.org/abs/1411.3441}
  {arXiv:1411.3441 [hep-ex]} \BibitemShut {NoStop}%
%%CITATION = ARXIV:1411.3441;%%
\bibitem [{\citenamefont {Aad}\ \emph {et~al.}(2015{\natexlab{b}})\citenamefont
  {Aad} \emph {et~al.}}]{Aad:2015wra}%
  \BibitemOpen
  \bibfield  {author} {\bibinfo {author} {\bibfnamefont {G.}~\bibnamefont
  {Aad}} \emph {et~al.} (\bibinfo {collaboration} {ATLAS}),\ }\href {\doibase
  10.1016/j.physletb.2015.03.054} {\bibfield  {journal} {\bibinfo  {journal}
  {Phys. Lett.}\ }\textbf {\bibinfo {volume} {B744}},\ \bibinfo {pages} {163}
  (\bibinfo {year} {2015}{\natexlab{b}})},\ \Eprint
  {http://arxiv.org/abs/1502.04478} {arXiv:1502.04478 [hep-ex]} \BibitemShut
  {NoStop}%
%%CITATION = ARXIV:1502.04478;%%
\bibitem [{\citenamefont {Aad}\ \emph {et~al.}(2016)\citenamefont {Aad} \emph
  {et~al.}}]{Khachatryan:2016vau}%
  \BibitemOpen
  \bibfield  {author} {\bibinfo {author} {\bibfnamefont {G.}~\bibnamefont
  {Aad}} \emph {et~al.} (\bibinfo {collaboration} {ATLAS, CMS}),\ }\href
  {\doibase 10.1007/JHEP08(2016)045} {\bibfield  {journal} {\bibinfo  {journal}
  {JHEP}\ }\textbf {\bibinfo {volume} {08}},\ \bibinfo {pages} {045} (\bibinfo
  {year} {2016})},\ \Eprint {http://arxiv.org/abs/1606.02266} {arXiv:1606.02266
  [hep-ex]} \BibitemShut {NoStop}%
%%CITATION = ARXIV:1606.02266;%%
\bibitem [{\citenamefont {Aaboud}\ \emph
  {et~al.}(2019{\natexlab{a}})\citenamefont {Aaboud} \emph
  {et~al.}}]{Aaboud:2018pen}%
  \BibitemOpen
  \bibfield  {author} {\bibinfo {author} {\bibfnamefont {M.}~\bibnamefont
  {Aaboud}} \emph {et~al.} (\bibinfo {collaboration} {ATLAS}),\ }\href
  {\doibase 10.1103/PhysRevD.99.072001} {\bibfield  {journal} {\bibinfo
  {journal} {Phys. Rev.}\ }\textbf {\bibinfo {volume} {D99}},\ \bibinfo {pages}
  {072001} (\bibinfo {year} {2019}{\natexlab{a}})},\ \Eprint
  {http://arxiv.org/abs/1811.08856} {arXiv:1811.08856 [hep-ex]} \BibitemShut
  {NoStop}%
%%CITATION = ARXIV:1811.08856;%%
\bibitem [{\citenamefont {Aaboud}\ \emph
  {et~al.}(2018{\natexlab{a}})\citenamefont {Aaboud} \emph
  {et~al.}}]{Aaboud:2018zhk}%
  \BibitemOpen
  \bibfield  {author} {\bibinfo {author} {\bibfnamefont {M.}~\bibnamefont
  {Aaboud}} \emph {et~al.} (\bibinfo {collaboration} {ATLAS}),\ }\href
  {\doibase 10.1016/j.physletb.2018.09.013} {\bibfield  {journal} {\bibinfo
  {journal} {Phys. Lett.}\ }\textbf {\bibinfo {volume} {B786}},\ \bibinfo
  {pages} {59} (\bibinfo {year} {2018}{\natexlab{a}})},\ \Eprint
  {http://arxiv.org/abs/1808.08238} {arXiv:1808.08238 [hep-ex]} \BibitemShut
  {NoStop}%
%%CITATION = ARXIV:1808.08238;%%
\bibitem [{\citenamefont {Noble}\ and\ \citenamefont
  {Perelstein}(2008)}]{Noble:2007kk}%
  \BibitemOpen
  \bibfield  {author} {\bibinfo {author} {\bibfnamefont {A.}~\bibnamefont
  {Noble}}\ and\ \bibinfo {author} {\bibfnamefont {M.}~\bibnamefont
  {Perelstein}},\ }\href {\doibase 10.1103/PhysRevD.78.063518} {\bibfield
  {journal} {\bibinfo  {journal} {Phys. Rev.}\ }\textbf {\bibinfo {volume}
  {D78}},\ \bibinfo {pages} {063518} (\bibinfo {year} {2008})},\ \Eprint
  {http://arxiv.org/abs/0711.3018} {arXiv:0711.3018 [hep-ph]} \BibitemShut
  {NoStop}%
%%CITATION = ARXIV:0711.3018;%%
\bibitem [{\citenamefont {Trodden}(1999)}]{Trodden:1998ym}%
  \BibitemOpen
  \bibfield  {author} {\bibinfo {author} {\bibfnamefont {M.}~\bibnamefont
  {Trodden}},\ }\href {\doibase 10.1103/RevModPhys.71.1463} {\bibfield
  {journal} {\bibinfo  {journal} {Rev. Mod. Phys.}\ }\textbf {\bibinfo {volume}
  {71}},\ \bibinfo {pages} {1463} (\bibinfo {year} {1999})},\ \Eprint
  {http://arxiv.org/abs/hep-ph/9803479} {arXiv:hep-ph/9803479 [hep-ph]}
  \BibitemShut {NoStop}%
%%CITATION = HEP-PH/9803479;%%
\bibitem [{\citenamefont {Morrissey}\ and\ \citenamefont
  {Ramsey-Musolf}(2012)}]{Morrissey:2012db}%
  \BibitemOpen
  \bibfield  {author} {\bibinfo {author} {\bibfnamefont {D.~E.}\ \bibnamefont
  {Morrissey}}\ and\ \bibinfo {author} {\bibfnamefont {M.~J.}\ \bibnamefont
  {Ramsey-Musolf}},\ }\href {\doibase 10.1088/1367-2630/14/12/125003}
  {\bibfield  {journal} {\bibinfo  {journal} {New J. Phys.}\ }\textbf {\bibinfo
  {volume} {14}},\ \bibinfo {pages} {125003} (\bibinfo {year} {2012})},\
  \Eprint {http://arxiv.org/abs/1206.2942} {arXiv:1206.2942 [hep-ph]}
  \BibitemShut {NoStop}%
%%CITATION = ARXIV:1206.2942;%%
\bibitem [{\citenamefont {Sirunyan}\ \emph
  {et~al.}(2019{\natexlab{a}})\citenamefont {Sirunyan} \emph
  {et~al.}}]{Sirunyan:2018iwt}%
  \BibitemOpen
  \bibfield  {author} {\bibinfo {author} {\bibfnamefont {A.~M.}\ \bibnamefont
  {Sirunyan}} \emph {et~al.} (\bibinfo {collaboration} {CMS}),\ }\href
  {\doibase 10.1016/j.physletb.2018.10.056} {\bibfield  {journal} {\bibinfo
  {journal} {Phys. Lett.}\ }\textbf {\bibinfo {volume} {B788}},\ \bibinfo
  {pages} {7} (\bibinfo {year} {2019}{\natexlab{a}})},\ \Eprint
  {http://arxiv.org/abs/1806.00408} {arXiv:1806.00408 [hep-ex]} \BibitemShut
  {NoStop}%
%%CITATION = ARXIV:1806.00408;%%
\bibitem [{\citenamefont {Aaboud}\ \emph
  {et~al.}(2018{\natexlab{b}})\citenamefont {Aaboud} \emph
  {et~al.}}]{Aaboud:2018ftw}%
  \BibitemOpen
  \bibfield  {author} {\bibinfo {author} {\bibfnamefont {M.}~\bibnamefont
  {Aaboud}} \emph {et~al.} (\bibinfo {collaboration} {ATLAS}),\ }\href
  {\doibase 10.1007/JHEP11(2018)040} {\bibfield  {journal} {\bibinfo  {journal}
  {JHEP}\ }\textbf {\bibinfo {volume} {11}},\ \bibinfo {pages} {040} (\bibinfo
  {year} {2018}{\natexlab{b}})},\ \Eprint {http://arxiv.org/abs/1807.04873}
  {arXiv:1807.04873 [hep-ex]} \BibitemShut {NoStop}%
%%CITATION = ARXIV:1807.04873;%%
\bibitem [{\citenamefont {Aaboud}\ \emph
  {et~al.}(2019{\natexlab{b}})\citenamefont {Aaboud} \emph
  {et~al.}}]{Aaboud:2018knk}%
  \BibitemOpen
  \bibfield  {author} {\bibinfo {author} {\bibfnamefont {M.}~\bibnamefont
  {Aaboud}} \emph {et~al.} (\bibinfo {collaboration} {ATLAS}),\ }\href
  {\doibase 10.1007/JHEP01(2019)030} {\bibfield  {journal} {\bibinfo  {journal}
  {JHEP}\ }\textbf {\bibinfo {volume} {01}},\ \bibinfo {pages} {030} (\bibinfo
  {year} {2019}{\natexlab{b}})},\ \Eprint {http://arxiv.org/abs/1804.06174}
  {arXiv:1804.06174 [hep-ex]} \BibitemShut {NoStop}%
%%CITATION = ARXIV:1804.06174;%%
\bibitem [{\citenamefont {Sirunyan}\ \emph
  {et~al.}(2019{\natexlab{b}})\citenamefont {Sirunyan} \emph
  {et~al.}}]{Sirunyan:2018two}%
  \BibitemOpen
  \bibfield  {author} {\bibinfo {author} {\bibfnamefont {A.~M.}\ \bibnamefont
  {Sirunyan}} \emph {et~al.} (\bibinfo {collaboration} {CMS}),\ }\href
  {\doibase 10.1103/PhysRevLett.122.121803} {\bibfield  {journal} {\bibinfo
  {journal} {Phys. Rev. Lett.}\ }\textbf {\bibinfo {volume} {122}},\ \bibinfo
  {pages} {121803} (\bibinfo {year} {2019}{\natexlab{b}})},\ \Eprint
  {http://arxiv.org/abs/1811.09689} {arXiv:1811.09689 [hep-ex]} \BibitemShut
  {NoStop}%
%%CITATION = ARXIV:1811.09689;%%
\bibitem [{\citenamefont {Aad}\ \emph {et~al.}(2019)\citenamefont {Aad} \emph
  {et~al.}}]{Aad:2019uzh}%
  \BibitemOpen
  \bibfield  {author} {\bibinfo {author} {\bibfnamefont {G.}~\bibnamefont
  {Aad}} \emph {et~al.} (\bibinfo {collaboration} {ATLAS}),\ }\href@noop {} {\
  (\bibinfo {year} {2019})},\ \Eprint {http://arxiv.org/abs/1906.02025}
  {arXiv:1906.02025 [hep-ex]} \BibitemShut {NoStop}%
%%CITATION = ARXIV:1906.02025;%%
\bibitem [{\citenamefont {Baur}\ \emph {et~al.}(2002)\citenamefont {Baur},
  \citenamefont {Plehn},\ and\ \citenamefont {Rainwater}}]{Baur:2002rb}%
  \BibitemOpen
  \bibfield  {author} {\bibinfo {author} {\bibfnamefont {U.}~\bibnamefont
  {Baur}}, \bibinfo {author} {\bibfnamefont {T.}~\bibnamefont {Plehn}}, \ and\
  \bibinfo {author} {\bibfnamefont {D.~L.}\ \bibnamefont {Rainwater}},\ }\href
  {\doibase 10.1103/PhysRevLett.89.151801} {\bibfield  {journal} {\bibinfo
  {journal} {Phys. Rev. Lett.}\ }\textbf {\bibinfo {volume} {89}},\ \bibinfo
  {pages} {151801} (\bibinfo {year} {2002})},\ \Eprint
  {http://arxiv.org/abs/hep-ph/0206024} {arXiv:hep-ph/0206024 [hep-ph]}
  \BibitemShut {NoStop}%
%%CITATION = HEP-PH/0206024;%%
\bibitem [{\citenamefont {Baur}\ \emph {et~al.}(2003)\citenamefont {Baur},
  \citenamefont {Plehn},\ and\ \citenamefont {Rainwater}}]{Baur:2002qd}%
  \BibitemOpen
  \bibfield  {author} {\bibinfo {author} {\bibfnamefont {U.}~\bibnamefont
  {Baur}}, \bibinfo {author} {\bibfnamefont {T.}~\bibnamefont {Plehn}}, \ and\
  \bibinfo {author} {\bibfnamefont {D.~L.}\ \bibnamefont {Rainwater}},\ }\href
  {\doibase 10.1103/PhysRevD.67.033003} {\bibfield  {journal} {\bibinfo
  {journal} {Phys. Rev.}\ }\textbf {\bibinfo {volume} {D67}},\ \bibinfo {pages}
  {033003} (\bibinfo {year} {2003})},\ \Eprint
  {http://arxiv.org/abs/hep-ph/0211224} {arXiv:hep-ph/0211224 [hep-ph]}
  \BibitemShut {NoStop}%
%%CITATION = HEP-PH/0211224;%%
\bibitem [{\citenamefont {Baur}\ \emph {et~al.}(2004)\citenamefont {Baur},
  \citenamefont {Plehn},\ and\ \citenamefont {Rainwater}}]{Baur:2003gp}%
  \BibitemOpen
  \bibfield  {author} {\bibinfo {author} {\bibfnamefont {U.}~\bibnamefont
  {Baur}}, \bibinfo {author} {\bibfnamefont {T.}~\bibnamefont {Plehn}}, \ and\
  \bibinfo {author} {\bibfnamefont {D.~L.}\ \bibnamefont {Rainwater}},\ }\href
  {\doibase 10.1103/PhysRevD.69.053004} {\bibfield  {journal} {\bibinfo
  {journal} {Phys. Rev.}\ }\textbf {\bibinfo {volume} {D69}},\ \bibinfo {pages}
  {053004} (\bibinfo {year} {2004})},\ \Eprint
  {http://arxiv.org/abs/hep-ph/0310056} {arXiv:hep-ph/0310056 [hep-ph]}
  \BibitemShut {NoStop}%
%%CITATION = HEP-PH/0310056;%%
\bibitem [{\citenamefont {Moretti}\ \emph {et~al.}(2005)\citenamefont
  {Moretti}, \citenamefont {Moretti}, \citenamefont {Piccinini}, \citenamefont
  {Pittau},\ and\ \citenamefont {Polosa}}]{Moretti:2004wa}%
  \BibitemOpen
  \bibfield  {author} {\bibinfo {author} {\bibfnamefont {M.}~\bibnamefont
  {Moretti}}, \bibinfo {author} {\bibfnamefont {S.}~\bibnamefont {Moretti}},
  \bibinfo {author} {\bibfnamefont {F.}~\bibnamefont {Piccinini}}, \bibinfo
  {author} {\bibfnamefont {R.}~\bibnamefont {Pittau}}, \ and\ \bibinfo {author}
  {\bibfnamefont {A.~D.}\ \bibnamefont {Polosa}},\ }\href {\doibase
  10.1088/1126-6708/2005/02/024} {\bibfield  {journal} {\bibinfo  {journal}
  {JHEP}\ }\textbf {\bibinfo {volume} {02}},\ \bibinfo {pages} {024} (\bibinfo
  {year} {2005})},\ \Eprint {http://arxiv.org/abs/hep-ph/0410334}
  {arXiv:hep-ph/0410334 [hep-ph]} \BibitemShut {NoStop}%
%%CITATION = HEP-PH/0410334;%%
\bibitem [{\citenamefont {Dolan}\ \emph {et~al.}(2012)\citenamefont {Dolan},
  \citenamefont {Englert},\ and\ \citenamefont {Spannowsky}}]{Dolan:2012rv}%
  \BibitemOpen
  \bibfield  {author} {\bibinfo {author} {\bibfnamefont {M.~J.}\ \bibnamefont
  {Dolan}}, \bibinfo {author} {\bibfnamefont {C.}~\bibnamefont {Englert}}, \
  and\ \bibinfo {author} {\bibfnamefont {M.}~\bibnamefont {Spannowsky}},\
  }\href {\doibase 10.1007/JHEP10(2012)112} {\bibfield  {journal} {\bibinfo
  {journal} {JHEP}\ }\textbf {\bibinfo {volume} {10}},\ \bibinfo {pages} {112}
  (\bibinfo {year} {2012})},\ \Eprint {http://arxiv.org/abs/1206.5001}
  {arXiv:1206.5001 [hep-ph]} \BibitemShut {NoStop}%
%%CITATION = ARXIV:1206.5001;%%
\bibitem [{\citenamefont {Baglio}\ \emph {et~al.}(2013)\citenamefont {Baglio},
  \citenamefont {Djouadi}, \citenamefont {Gr{\"o}ber}, \citenamefont
  {M{\"u}hlleitner}, \citenamefont {Quevillon},\ and\ \citenamefont
  {Spira}}]{Baglio:2012np}%
  \BibitemOpen
  \bibfield  {author} {\bibinfo {author} {\bibfnamefont {J.}~\bibnamefont
  {Baglio}}, \bibinfo {author} {\bibfnamefont {A.}~\bibnamefont {Djouadi}},
  \bibinfo {author} {\bibfnamefont {R.}~\bibnamefont {Gr{\"o}ber}}, \bibinfo
  {author} {\bibfnamefont {M.~M.}\ \bibnamefont {M{\"u}hlleitner}}, \bibinfo
  {author} {\bibfnamefont {J.}~\bibnamefont {Quevillon}}, \ and\ \bibinfo
  {author} {\bibfnamefont {M.}~\bibnamefont {Spira}},\ }\href {\doibase
  10.1007/JHEP04(2013)151} {\bibfield  {journal} {\bibinfo  {journal} {JHEP}\
  }\textbf {\bibinfo {volume} {04}},\ \bibinfo {pages} {151} (\bibinfo {year}
  {2013})},\ \Eprint {http://arxiv.org/abs/1212.5581} {arXiv:1212.5581
  [hep-ph]} \BibitemShut {NoStop}%
%%CITATION = ARXIV:1212.5581;%%
\bibitem [{\citenamefont {Goertz}\ \emph {et~al.}(2013)\citenamefont {Goertz},
  \citenamefont {Papaefstathiou}, \citenamefont {Yang},\ and\ \citenamefont
  {Zurita}}]{Goertz:2013eka}%
  \BibitemOpen
  \bibfield  {author} {\bibinfo {author} {\bibfnamefont {F.}~\bibnamefont
  {Goertz}}, \bibinfo {author} {\bibfnamefont {A.}~\bibnamefont
  {Papaefstathiou}}, \bibinfo {author} {\bibfnamefont {L.~L.}\ \bibnamefont
  {Yang}}, \ and\ \bibinfo {author} {\bibfnamefont {J.}~\bibnamefont
  {Zurita}},\ }in\ \href@noop {} {\emph {\bibinfo {booktitle} {{25th Rencontres
  de Blois on Particle Physics and Cosmology Blois, France, May 26-31,
  2013}}}}\ (\bibinfo {year} {2013})\ \Eprint {http://arxiv.org/abs/1309.3805}
  {arXiv:1309.3805 [hep-ph]} \BibitemShut {NoStop}%
%%CITATION = ARXIV:1309.3805;%%
\bibitem [{\citenamefont {Frederix}\ \emph {et~al.}(2014)\citenamefont
  {Frederix}, \citenamefont {Frixione}, \citenamefont {Hirschi}, \citenamefont
  {Maltoni}, \citenamefont {Mattelaer}, \citenamefont {Torrielli},
  \citenamefont {Vryonidou},\ and\ \citenamefont {Zaro}}]{Frederix:2014hta}%
  \BibitemOpen
  \bibfield  {author} {\bibinfo {author} {\bibfnamefont {R.}~\bibnamefont
  {Frederix}}, \bibinfo {author} {\bibfnamefont {S.}~\bibnamefont {Frixione}},
  \bibinfo {author} {\bibfnamefont {V.}~\bibnamefont {Hirschi}}, \bibinfo
  {author} {\bibfnamefont {F.}~\bibnamefont {Maltoni}}, \bibinfo {author}
  {\bibfnamefont {O.}~\bibnamefont {Mattelaer}}, \bibinfo {author}
  {\bibfnamefont {P.}~\bibnamefont {Torrielli}}, \bibinfo {author}
  {\bibfnamefont {E.}~\bibnamefont {Vryonidou}}, \ and\ \bibinfo {author}
  {\bibfnamefont {M.}~\bibnamefont {Zaro}},\ }\href {\doibase
  10.1016/j.physletb.2014.03.026} {\bibfield  {journal} {\bibinfo  {journal}
  {Phys. Lett.}\ }\textbf {\bibinfo {volume} {B732}},\ \bibinfo {pages} {142}
  (\bibinfo {year} {2014})},\ \Eprint {http://arxiv.org/abs/1401.7340}
  {arXiv:1401.7340 [hep-ph]} \BibitemShut {NoStop}%
%%CITATION = ARXIV:1401.7340;%%
\bibitem [{\citenamefont {Cao}\ \emph {et~al.}(2017)\citenamefont {Cao},
  \citenamefont {Liu},\ and\ \citenamefont {Yan}}]{Cao:2015oxx}%
  \BibitemOpen
  \bibfield  {author} {\bibinfo {author} {\bibfnamefont {Q.-H.}\ \bibnamefont
  {Cao}}, \bibinfo {author} {\bibfnamefont {Y.}~\bibnamefont {Liu}}, \ and\
  \bibinfo {author} {\bibfnamefont {B.}~\bibnamefont {Yan}},\ }\href {\doibase
  10.1103/PhysRevD.95.073006} {\bibfield  {journal} {\bibinfo  {journal} {Phys.
  Rev.}\ }\textbf {\bibinfo {volume} {D95}},\ \bibinfo {pages} {073006}
  (\bibinfo {year} {2017})},\ \Eprint {http://arxiv.org/abs/1511.03311}
  {arXiv:1511.03311 [hep-ph]} \BibitemShut {NoStop}%
%%CITATION = ARXIV:1511.03311;%%
\bibitem [{\citenamefont {Gouzevitch}\ \emph {et~al.}(2013)\citenamefont
  {Gouzevitch}, \citenamefont {Oliveira}, \citenamefont {Rojo}, \citenamefont
  {Rosenfeld}, \citenamefont {Salam},\ and\ \citenamefont
  {Sanz}}]{Gouzevitch:2013qca}%
  \BibitemOpen
  \bibfield  {author} {\bibinfo {author} {\bibfnamefont {M.}~\bibnamefont
  {Gouzevitch}}, \bibinfo {author} {\bibfnamefont {A.}~\bibnamefont
  {Oliveira}}, \bibinfo {author} {\bibfnamefont {J.}~\bibnamefont {Rojo}},
  \bibinfo {author} {\bibfnamefont {R.}~\bibnamefont {Rosenfeld}}, \bibinfo
  {author} {\bibfnamefont {G.~P.}\ \bibnamefont {Salam}}, \ and\ \bibinfo
  {author} {\bibfnamefont {V.}~\bibnamefont {Sanz}},\ }\href {\doibase
  10.1007/JHEP07(2013)148} {\bibfield  {journal} {\bibinfo  {journal} {JHEP}\
  }\textbf {\bibinfo {volume} {07}},\ \bibinfo {pages} {148} (\bibinfo {year}
  {2013})},\ \Eprint {http://arxiv.org/abs/1303.6636} {arXiv:1303.6636
  [hep-ph]} \BibitemShut {NoStop}%
%%CITATION = ARXIV:1303.6636;%%
\bibitem [{\citenamefont {Behr}\ \emph {et~al.}(2016)\citenamefont {Behr},
  \citenamefont {Bortoletto}, \citenamefont {Frost}, \citenamefont {Hartland},
  \citenamefont {Issever},\ and\ \citenamefont {Rojo}}]{Behr:2015oqq}%
  \BibitemOpen
  \bibfield  {author} {\bibinfo {author} {\bibfnamefont {J.~K.}\ \bibnamefont
  {Behr}}, \bibinfo {author} {\bibfnamefont {D.}~\bibnamefont {Bortoletto}},
  \bibinfo {author} {\bibfnamefont {J.~A.}\ \bibnamefont {Frost}}, \bibinfo
  {author} {\bibfnamefont {N.~P.}\ \bibnamefont {Hartland}}, \bibinfo {author}
  {\bibfnamefont {C.}~\bibnamefont {Issever}}, \ and\ \bibinfo {author}
  {\bibfnamefont {J.}~\bibnamefont {Rojo}},\ }\href {\doibase
  10.1140/epjc/s10052-016-4215-5} {\bibfield  {journal} {\bibinfo  {journal}
  {Eur. Phys. J.}\ }\textbf {\bibinfo {volume} {C76}},\ \bibinfo {pages} {386}
  (\bibinfo {year} {2016})},\ \Eprint {http://arxiv.org/abs/1512.08928}
  {arXiv:1512.08928 [hep-ph]} \BibitemShut {NoStop}%
%%CITATION = ARXIV:1512.08928;%%
\bibitem [{\citenamefont {Bishara}\ \emph {et~al.}(2017)\citenamefont
  {Bishara}, \citenamefont {Contino},\ and\ \citenamefont
  {Rojo}}]{Bishara:2016kjn}%
  \BibitemOpen
  \bibfield  {author} {\bibinfo {author} {\bibfnamefont {F.}~\bibnamefont
  {Bishara}}, \bibinfo {author} {\bibfnamefont {R.}~\bibnamefont {Contino}}, \
  and\ \bibinfo {author} {\bibfnamefont {J.}~\bibnamefont {Rojo}},\ }\href
  {\doibase 10.1140/epjc/s10052-017-5037-9} {\bibfield  {journal} {\bibinfo
  {journal} {Eur. Phys. J.}\ }\textbf {\bibinfo {volume} {C77}},\ \bibinfo
  {pages} {481} (\bibinfo {year} {2017})},\ \Eprint
  {http://arxiv.org/abs/1611.03860} {arXiv:1611.03860 [hep-ph]} \BibitemShut
  {NoStop}%
%%CITATION = ARXIV:1611.03860;%%
\bibitem [{\citenamefont {Di~Vita}\ \emph {et~al.}(2017)\citenamefont
  {Di~Vita}, \citenamefont {Grojean}, \citenamefont {Panico}, \citenamefont
  {Riembau},\ and\ \citenamefont {Vantalon}}]{DiVita:2017eyz}%
  \BibitemOpen
  \bibfield  {author} {\bibinfo {author} {\bibfnamefont {S.}~\bibnamefont
  {Di~Vita}}, \bibinfo {author} {\bibfnamefont {C.}~\bibnamefont {Grojean}},
  \bibinfo {author} {\bibfnamefont {G.}~\bibnamefont {Panico}}, \bibinfo
  {author} {\bibfnamefont {M.}~\bibnamefont {Riembau}}, \ and\ \bibinfo
  {author} {\bibfnamefont {T.}~\bibnamefont {Vantalon}},\ }\href {\doibase
  10.1007/JHEP09(2017)069} {\bibfield  {journal} {\bibinfo  {journal} {JHEP}\
  }\textbf {\bibinfo {volume} {09}},\ \bibinfo {pages} {069} (\bibinfo {year}
  {2017})},\ \Eprint {http://arxiv.org/abs/1704.01953} {arXiv:1704.01953
  [hep-ph]} \BibitemShut {NoStop}%
%%CITATION = ARXIV:1704.01953;%%
\bibitem [{\citenamefont {Cepeda}\ \emph {et~al.}(2019)\citenamefont {Cepeda}
  \emph {et~al.}}]{Cepeda:2019klc}%
  \BibitemOpen
  \bibfield  {author} {\bibinfo {author} {\bibfnamefont {M.}~\bibnamefont
  {Cepeda}} \emph {et~al.} (\bibinfo {collaboration} {HL/HE WG2 group}),\
  }\href@noop {} {\  (\bibinfo {year} {2019})},\ \Eprint
  {http://arxiv.org/abs/1902.00134} {arXiv:1902.00134 [hep-ph]} \BibitemShut
  {NoStop}%
%%CITATION = ARXIV:1902.00134;%%
\bibitem [{\citenamefont {Baer}\ \emph {et~al.}(2013)\citenamefont {Baer},
  \citenamefont {Barklow}, \citenamefont {Fujii}, \citenamefont {Gao},
  \citenamefont {Hoang}, \citenamefont {Kanemura}, \citenamefont {List},
  \citenamefont {Logan}, \citenamefont {Nomerotski}, \citenamefont {Perelstein}
  \emph {et~al.}}]{Baer:2013cma}%
  \BibitemOpen
  \bibfield  {author} {\bibinfo {author} {\bibfnamefont {H.}~\bibnamefont
  {Baer}}, \bibinfo {author} {\bibfnamefont {T.}~\bibnamefont {Barklow}},
  \bibinfo {author} {\bibfnamefont {K.}~\bibnamefont {Fujii}}, \bibinfo
  {author} {\bibfnamefont {Y.}~\bibnamefont {Gao}}, \bibinfo {author}
  {\bibfnamefont {A.}~\bibnamefont {Hoang}}, \bibinfo {author} {\bibfnamefont
  {S.}~\bibnamefont {Kanemura}}, \bibinfo {author} {\bibfnamefont
  {J.}~\bibnamefont {List}}, \bibinfo {author} {\bibfnamefont {H.~E.}\
  \bibnamefont {Logan}}, \bibinfo {author} {\bibfnamefont {A.}~\bibnamefont
  {Nomerotski}}, \bibinfo {author} {\bibfnamefont {M.}~\bibnamefont
  {Perelstein}},  \emph {et~al.},\ }\href@noop {} {\  (\bibinfo {year}
  {2013})},\ \Eprint {http://arxiv.org/abs/1306.6352} {arXiv:1306.6352
  [hep-ph]} \BibitemShut {NoStop}%
%%CITATION = ARXIV:1306.6352;%%
\bibitem [{\citenamefont {Asner}\ \emph {et~al.}(2013)\citenamefont {Asner}
  \emph {et~al.}}]{Asner:2013psa}%
  \BibitemOpen
  \bibfield  {author} {\bibinfo {author} {\bibfnamefont {D.~M.}\ \bibnamefont
  {Asner}} \emph {et~al.},\ }in\ \href@noop {} {\emph {\bibinfo {booktitle}
  {{Proceedings, 2013 Community Summer Study on the Future of U.S. Particle
  Physics: Snowmass on the Mississippi (CSS2013): Minneapolis, MN, USA, July
  29-August 6, 2013}}}}\ (\bibinfo {year} {2013})\ \Eprint
  {http://arxiv.org/abs/1310.0763} {arXiv:1310.0763 [hep-ph]} \BibitemShut
  {NoStop}%
%%CITATION = ARXIV:1310.0763;%%
\bibitem [{\citenamefont {Di~Vita}\ \emph {et~al.}(2018)\citenamefont
  {Di~Vita}, \citenamefont {Durieux}, \citenamefont {Grojean}, \citenamefont
  {Gu}, \citenamefont {Liu}, \citenamefont {Panico}, \citenamefont {Riembau},\
  and\ \citenamefont {Vantalon}}]{DiVita:2017vrr}%
  \BibitemOpen
  \bibfield  {author} {\bibinfo {author} {\bibfnamefont {S.}~\bibnamefont
  {Di~Vita}}, \bibinfo {author} {\bibfnamefont {G.}~\bibnamefont {Durieux}},
  \bibinfo {author} {\bibfnamefont {C.}~\bibnamefont {Grojean}}, \bibinfo
  {author} {\bibfnamefont {J.}~\bibnamefont {Gu}}, \bibinfo {author}
  {\bibfnamefont {Z.}~\bibnamefont {Liu}}, \bibinfo {author} {\bibfnamefont
  {G.}~\bibnamefont {Panico}}, \bibinfo {author} {\bibfnamefont
  {M.}~\bibnamefont {Riembau}}, \ and\ \bibinfo {author} {\bibfnamefont
  {T.}~\bibnamefont {Vantalon}},\ }\href {\doibase 10.1007/JHEP02(2018)178}
  {\bibfield  {journal} {\bibinfo  {journal} {JHEP}\ }\textbf {\bibinfo
  {volume} {02}},\ \bibinfo {pages} {178} (\bibinfo {year} {2018})},\ \Eprint
  {http://arxiv.org/abs/1711.03978} {arXiv:1711.03978 [hep-ph]} \BibitemShut
  {NoStop}%
%%CITATION = ARXIV:1711.03978;%%
\bibitem [{\citenamefont {Maltoni}\ \emph {et~al.}(2018)\citenamefont
  {Maltoni}, \citenamefont {Pagani},\ and\ \citenamefont
  {Zhao}}]{Maltoni:2018ttu}%
  \BibitemOpen
  \bibfield  {author} {\bibinfo {author} {\bibfnamefont {F.}~\bibnamefont
  {Maltoni}}, \bibinfo {author} {\bibfnamefont {D.}~\bibnamefont {Pagani}}, \
  and\ \bibinfo {author} {\bibfnamefont {X.}~\bibnamefont {Zhao}},\ }\href
  {\doibase 10.1007/JHEP07(2018)087} {\bibfield  {journal} {\bibinfo  {journal}
  {JHEP}\ }\textbf {\bibinfo {volume} {07}},\ \bibinfo {pages} {087} (\bibinfo
  {year} {2018})},\ \Eprint {http://arxiv.org/abs/1802.07616} {arXiv:1802.07616
  [hep-ph]} \BibitemShut {NoStop}%
%%CITATION = ARXIV:1802.07616;%%
\bibitem [{\citenamefont {Yao}(2013)}]{Yao:2013ika}%
  \BibitemOpen
  \bibfield  {author} {\bibinfo {author} {\bibfnamefont {W.}~\bibnamefont
  {Yao}},\ }in\ \href
  {http://www.slac.stanford.edu/econf/C1307292/docs/submittedArxivFiles/1308.6302.pdf}
  {\emph {\bibinfo {booktitle} {{Proceedings, 2013 Community Summer Study on
  the Future of U.S. Particle Physics: Snowmass on the Mississippi (CSS2013):
  Minneapolis, MN, USA, July 29-August 6, 2013}}}}\ (\bibinfo {year} {2013})\
  \Eprint {http://arxiv.org/abs/1308.6302} {arXiv:1308.6302 [hep-ph]}
  \BibitemShut {NoStop}%
%%CITATION = ARXIV:1308.6302;%%
\bibitem [{\citenamefont {Barr}\ \emph {et~al.}(2015)\citenamefont {Barr},
  \citenamefont {Dolan}, \citenamefont {Englert}, \citenamefont {Ferreira~de
  Lima},\ and\ \citenamefont {Spannowsky}}]{Barr:2014sga}%
  \BibitemOpen
  \bibfield  {author} {\bibinfo {author} {\bibfnamefont {A.~J.}\ \bibnamefont
  {Barr}}, \bibinfo {author} {\bibfnamefont {M.~J.}\ \bibnamefont {Dolan}},
  \bibinfo {author} {\bibfnamefont {C.}~\bibnamefont {Englert}}, \bibinfo
  {author} {\bibfnamefont {D.~E.}\ \bibnamefont {Ferreira~de Lima}}, \ and\
  \bibinfo {author} {\bibfnamefont {M.}~\bibnamefont {Spannowsky}},\ }\href
  {\doibase 10.1007/JHEP02(2015)016} {\bibfield  {journal} {\bibinfo  {journal}
  {JHEP}\ }\textbf {\bibinfo {volume} {02}},\ \bibinfo {pages} {016} (\bibinfo
  {year} {2015})},\ \Eprint {http://arxiv.org/abs/1412.7154} {arXiv:1412.7154
  [hep-ph]} \BibitemShut {NoStop}%
%%CITATION = ARXIV:1412.7154;%%
\bibitem [{\citenamefont {Azatov}\ \emph {et~al.}(2015)\citenamefont {Azatov},
  \citenamefont {Contino}, \citenamefont {Panico},\ and\ \citenamefont
  {Son}}]{Azatov:2015oxa}%
  \BibitemOpen
  \bibfield  {author} {\bibinfo {author} {\bibfnamefont {A.}~\bibnamefont
  {Azatov}}, \bibinfo {author} {\bibfnamefont {R.}~\bibnamefont {Contino}},
  \bibinfo {author} {\bibfnamefont {G.}~\bibnamefont {Panico}}, \ and\ \bibinfo
  {author} {\bibfnamefont {M.}~\bibnamefont {Son}},\ }\href {\doibase
  10.1103/PhysRevD.92.035001} {\bibfield  {journal} {\bibinfo  {journal} {Phys.
  Rev.}\ }\textbf {\bibinfo {volume} {D92}},\ \bibinfo {pages} {035001}
  (\bibinfo {year} {2015})},\ \Eprint {http://arxiv.org/abs/1502.00539}
  {arXiv:1502.00539 [hep-ph]} \BibitemShut {NoStop}%
%%CITATION = ARXIV:1502.00539;%%
\bibitem [{\citenamefont {He}\ \emph {et~al.}(2016)\citenamefont {He},
  \citenamefont {Ren},\ and\ \citenamefont {Yao}}]{He:2015spf}%
  \BibitemOpen
  \bibfield  {author} {\bibinfo {author} {\bibfnamefont {H.-J.}\ \bibnamefont
  {He}}, \bibinfo {author} {\bibfnamefont {J.}~\bibnamefont {Ren}}, \ and\
  \bibinfo {author} {\bibfnamefont {W.}~\bibnamefont {Yao}},\ }\href {\doibase
  10.1103/PhysRevD.93.015003} {\bibfield  {journal} {\bibinfo  {journal} {Phys.
  Rev.}\ }\textbf {\bibinfo {volume} {D93}},\ \bibinfo {pages} {015003}
  (\bibinfo {year} {2016})},\ \Eprint {http://arxiv.org/abs/1506.03302}
  {arXiv:1506.03302 [hep-ph]} \BibitemShut {NoStop}%
%%CITATION = ARXIV:1506.03302;%%
\bibitem [{\citenamefont {Chen}\ \emph {et~al.}(2016)\citenamefont {Chen},
  \citenamefont {Yan}, \citenamefont {Zhao}, \citenamefont {Zhong},\ and\
  \citenamefont {Zhao}}]{Chen:2015gva}%
  \BibitemOpen
  \bibfield  {author} {\bibinfo {author} {\bibfnamefont {C.-Y.}\ \bibnamefont
  {Chen}}, \bibinfo {author} {\bibfnamefont {Q.-S.}\ \bibnamefont {Yan}},
  \bibinfo {author} {\bibfnamefont {X.}~\bibnamefont {Zhao}}, \bibinfo {author}
  {\bibfnamefont {Y.-M.}\ \bibnamefont {Zhong}}, \ and\ \bibinfo {author}
  {\bibfnamefont {Z.}~\bibnamefont {Zhao}},\ }\href {\doibase
  10.1103/PhysRevD.93.013007} {\bibfield  {journal} {\bibinfo  {journal} {Phys.
  Rev.}\ }\textbf {\bibinfo {volume} {D93}},\ \bibinfo {pages} {013007}
  (\bibinfo {year} {2016})},\ \Eprint {http://arxiv.org/abs/1510.04013}
  {arXiv:1510.04013 [hep-ph]} \BibitemShut {NoStop}%
%%CITATION = ARXIV:1510.04013;%%
\bibitem [{\citenamefont {Contino}\ \emph {et~al.}(2017)\citenamefont {Contino}
  \emph {et~al.}}]{Contino:2016spe}%
  \BibitemOpen
  \bibfield  {author} {\bibinfo {author} {\bibfnamefont {R.}~\bibnamefont
  {Contino}} \emph {et~al.},\ }\href {\doibase 10.23731/CYRM-2017-003.255}
  {\bibfield  {journal} {\bibinfo  {journal} {CERN Yellow Rep.}\ ,\ \bibinfo
  {pages} {255}} (\bibinfo {year} {2017})},\ \Eprint
  {http://arxiv.org/abs/1606.09408} {arXiv:1606.09408 [hep-ph]} \BibitemShut
  {NoStop}%
%%CITATION = ARXIV:1606.09408;%%
\bibitem [{\citenamefont {Banerjee}\ \emph {et~al.}(2018)\citenamefont
  {Banerjee}, \citenamefont {Englert}, \citenamefont {Mangano}, \citenamefont
  {Selvaggi},\ and\ \citenamefont {Spannowsky}}]{Banerjee:2018yxy}%
  \BibitemOpen
  \bibfield  {author} {\bibinfo {author} {\bibfnamefont {S.}~\bibnamefont
  {Banerjee}}, \bibinfo {author} {\bibfnamefont {C.}~\bibnamefont {Englert}},
  \bibinfo {author} {\bibfnamefont {M.~L.}\ \bibnamefont {Mangano}}, \bibinfo
  {author} {\bibfnamefont {M.}~\bibnamefont {Selvaggi}}, \ and\ \bibinfo
  {author} {\bibfnamefont {M.}~\bibnamefont {Spannowsky}},\ }\href {\doibase
  10.1140/epjc/s10052-018-5788-y} {\bibfield  {journal} {\bibinfo  {journal}
  {Eur. Phys. J.}\ }\textbf {\bibinfo {volume} {C78}},\ \bibinfo {pages} {322}
  (\bibinfo {year} {2018})},\ \Eprint {http://arxiv.org/abs/1802.01607}
  {arXiv:1802.01607 [hep-ph]} \BibitemShut {NoStop}%
%%CITATION = ARXIV:1802.01607;%%
\bibitem [{\citenamefont {Chang}\ \emph {et~al.}(2018)\citenamefont {Chang},
  \citenamefont {Cheung}, \citenamefont {Lee}, \citenamefont {Lu},\ and\
  \citenamefont {Park}}]{Chang:2018uwu}%
  \BibitemOpen
  \bibfield  {author} {\bibinfo {author} {\bibfnamefont {J.}~\bibnamefont
  {Chang}}, \bibinfo {author} {\bibfnamefont {K.}~\bibnamefont {Cheung}},
  \bibinfo {author} {\bibfnamefont {J.~S.}\ \bibnamefont {Lee}}, \bibinfo
  {author} {\bibfnamefont {C.-T.}\ \bibnamefont {Lu}}, \ and\ \bibinfo {author}
  {\bibfnamefont {J.}~\bibnamefont {Park}},\ }\href@noop {} {\  (\bibinfo
  {year} {2018})},\ \Eprint {http://arxiv.org/abs/1804.07130} {arXiv:1804.07130
  [hep-ph]} \BibitemShut {NoStop}%
%%CITATION = ARXIV:1804.07130;%%
\bibitem [{\citenamefont {Blondel}\ and\ \citenamefont
  {Janot}(2018)}]{Blondel:2018aan}%
  \BibitemOpen
  \bibfield  {author} {\bibinfo {author} {\bibfnamefont {A.}~\bibnamefont
  {Blondel}}\ and\ \bibinfo {author} {\bibfnamefont {P.}~\bibnamefont
  {Janot}},\ }\href@noop {} {\  (\bibinfo {year} {2018})},\ \Eprint
  {http://arxiv.org/abs/1809.10041} {arXiv:1809.10041 [hep-ph]} \BibitemShut
  {NoStop}%
%%CITATION = ARXIV:1809.10041;%%
\bibitem [{\citenamefont {Kim}\ \emph {et~al.}(2018)\citenamefont {Kim},
  \citenamefont {Sakaki},\ and\ \citenamefont {Son}}]{Kim:2018uty}%
  \BibitemOpen
  \bibfield  {author} {\bibinfo {author} {\bibfnamefont {J.~H.}\ \bibnamefont
  {Kim}}, \bibinfo {author} {\bibfnamefont {Y.}~\bibnamefont {Sakaki}}, \ and\
  \bibinfo {author} {\bibfnamefont {M.}~\bibnamefont {Son}},\ }\href {\doibase
  10.1103/PhysRevD.98.015016} {\bibfield  {journal} {\bibinfo  {journal} {Phys.
  Rev.}\ }\textbf {\bibinfo {volume} {D98}},\ \bibinfo {pages} {015016}
  (\bibinfo {year} {2018})},\ \Eprint {http://arxiv.org/abs/1801.06093}
  {arXiv:1801.06093 [hep-ph]} \BibitemShut {NoStop}%
%%CITATION = ARXIV:1801.06093;%%
\bibitem [{\citenamefont {Kim}\ \emph {et~al.}(2019{\natexlab{a}})\citenamefont
  {Kim}, \citenamefont {Kong}, \citenamefont {Matchev},\ and\ \citenamefont
  {Park}}]{Kim:2018cxf}%
  \BibitemOpen
  \bibfield  {author} {\bibinfo {author} {\bibfnamefont {J.~H.}\ \bibnamefont
  {Kim}}, \bibinfo {author} {\bibfnamefont {K.}~\bibnamefont {Kong}}, \bibinfo
  {author} {\bibfnamefont {K.~T.}\ \bibnamefont {Matchev}}, \ and\ \bibinfo
  {author} {\bibfnamefont {M.}~\bibnamefont {Park}},\ }\href {\doibase
  10.1103/PhysRevLett.122.091801} {\bibfield  {journal} {\bibinfo  {journal}
  {Phys. Rev. Lett.}\ }\textbf {\bibinfo {volume} {122}},\ \bibinfo {pages}
  {091801} (\bibinfo {year} {2019}{\natexlab{a}})},\ \Eprint
  {http://arxiv.org/abs/1807.11498} {arXiv:1807.11498 [hep-ph]} \BibitemShut
  {NoStop}%
%%CITATION = ARXIV:1807.11498;%%
\bibitem [{\citenamefont {Kim}\ \emph {et~al.}(2019{\natexlab{b}})\citenamefont
  {Kim}, \citenamefont {Kim}, \citenamefont {Kong}, \citenamefont {Matchev},\
  and\ \citenamefont {Park}}]{Kim:2019wns}%
  \BibitemOpen
  \bibfield  {author} {\bibinfo {author} {\bibfnamefont {J.~H.}\ \bibnamefont
  {Kim}}, \bibinfo {author} {\bibfnamefont {M.}~\bibnamefont {Kim}}, \bibinfo
  {author} {\bibfnamefont {K.}~\bibnamefont {Kong}}, \bibinfo {author}
  {\bibfnamefont {K.~T.}\ \bibnamefont {Matchev}}, \ and\ \bibinfo {author}
  {\bibfnamefont {M.}~\bibnamefont {Park}},\ }\href {\doibase
  10.1007/JHEP09(2019)047} {\bibfield  {journal} {\bibinfo  {journal} {JHEP}\
  }\textbf {\bibinfo {volume} {09}},\ \bibinfo {pages} {047} (\bibinfo {year}
  {2019}{\natexlab{b}})},\ \Eprint {http://arxiv.org/abs/1904.08549}
  {arXiv:1904.08549 [hep-ph]} \BibitemShut {NoStop}%
%%CITATION = ARXIV:1904.08549;%%
\bibitem [{\citenamefont {McCullough}(2014)}]{McCullough:2013rea}%
  \BibitemOpen
  \bibfield  {author} {\bibinfo {author} {\bibfnamefont {M.}~\bibnamefont
  {McCullough}},\ }\href {\doibase 10.1103/PhysRevD.90.015001,
  10.1103/PhysRevD.92.039903} {\bibfield  {journal} {\bibinfo  {journal} {Phys.
  Rev.}\ }\textbf {\bibinfo {volume} {D90}},\ \bibinfo {pages} {015001}
  (\bibinfo {year} {2014})},\ \bibinfo {note} {[Erratum: Phys.
  Rev.D92,no.3,039903(2015)]},\ \Eprint {http://arxiv.org/abs/1312.3322}
  {arXiv:1312.3322 [hep-ph]} \BibitemShut {NoStop}%
%%CITATION = ARXIV:1312.3322;%%
\bibitem [{\citenamefont {Shen}\ and\ \citenamefont
  {Zhu}(2015)}]{Shen:2015pha}%
  \BibitemOpen
  \bibfield  {author} {\bibinfo {author} {\bibfnamefont {C.}~\bibnamefont
  {Shen}}\ and\ \bibinfo {author} {\bibfnamefont {S.-h.}\ \bibnamefont {Zhu}},\
  }\href {\doibase 10.1103/PhysRevD.92.094001} {\bibfield  {journal} {\bibinfo
  {journal} {Phys. Rev.}\ }\textbf {\bibinfo {volume} {D92}},\ \bibinfo {pages}
  {094001} (\bibinfo {year} {2015})},\ \Eprint
  {http://arxiv.org/abs/1504.05626} {arXiv:1504.05626 [hep-ph]} \BibitemShut
  {NoStop}%
%%CITATION = ARXIV:1504.05626;%%
\bibitem [{\citenamefont {Degrassi}\ \emph {et~al.}(2016)\citenamefont
  {Degrassi}, \citenamefont {Giardino}, \citenamefont {Maltoni},\ and\
  \citenamefont {Pagani}}]{Degrassi:2016wml}%
  \BibitemOpen
  \bibfield  {author} {\bibinfo {author} {\bibfnamefont {G.}~\bibnamefont
  {Degrassi}}, \bibinfo {author} {\bibfnamefont {P.~P.}\ \bibnamefont
  {Giardino}}, \bibinfo {author} {\bibfnamefont {F.}~\bibnamefont {Maltoni}}, \
  and\ \bibinfo {author} {\bibfnamefont {D.}~\bibnamefont {Pagani}},\ }\href
  {\doibase 10.1007/JHEP12(2016)080} {\bibfield  {journal} {\bibinfo  {journal}
  {JHEP}\ }\textbf {\bibinfo {volume} {12}},\ \bibinfo {pages} {080} (\bibinfo
  {year} {2016})},\ \Eprint {http://arxiv.org/abs/1607.04251} {arXiv:1607.04251
  [hep-ph]} \BibitemShut {NoStop}%
%%CITATION = ARXIV:1607.04251;%%
\bibitem [{\citenamefont {Bizon}\ \emph {et~al.}(2017)\citenamefont {Bizon},
  \citenamefont {Gorbahn}, \citenamefont {Haisch},\ and\ \citenamefont
  {Zanderighi}}]{Bizon:2016wgr}%
  \BibitemOpen
  \bibfield  {author} {\bibinfo {author} {\bibfnamefont {W.}~\bibnamefont
  {Bizon}}, \bibinfo {author} {\bibfnamefont {M.}~\bibnamefont {Gorbahn}},
  \bibinfo {author} {\bibfnamefont {U.}~\bibnamefont {Haisch}}, \ and\ \bibinfo
  {author} {\bibfnamefont {G.}~\bibnamefont {Zanderighi}},\ }\href {\doibase
  10.1007/JHEP07(2017)083} {\bibfield  {journal} {\bibinfo  {journal} {JHEP}\
  }\textbf {\bibinfo {volume} {07}},\ \bibinfo {pages} {083} (\bibinfo {year}
  {2017})},\ \Eprint {http://arxiv.org/abs/1610.05771} {arXiv:1610.05771
  [hep-ph]} \BibitemShut {NoStop}%
%%CITATION = ARXIV:1610.05771;%%
\bibitem [{\citenamefont {Maltoni}\ \emph {et~al.}(2017)\citenamefont
  {Maltoni}, \citenamefont {Pagani}, \citenamefont {Shivaji},\ and\
  \citenamefont {Zhao}}]{Maltoni:2017ims}%
  \BibitemOpen
  \bibfield  {author} {\bibinfo {author} {\bibfnamefont {F.}~\bibnamefont
  {Maltoni}}, \bibinfo {author} {\bibfnamefont {D.}~\bibnamefont {Pagani}},
  \bibinfo {author} {\bibfnamefont {A.}~\bibnamefont {Shivaji}}, \ and\
  \bibinfo {author} {\bibfnamefont {X.}~\bibnamefont {Zhao}},\ }\href {\doibase
  10.1140/epjc/s10052-017-5410-8} {\bibfield  {journal} {\bibinfo  {journal}
  {Eur. Phys. J.}\ }\textbf {\bibinfo {volume} {C77}},\ \bibinfo {pages} {887}
  (\bibinfo {year} {2017})},\ \Eprint {http://arxiv.org/abs/1709.08649}
  {arXiv:1709.08649 [hep-ph]} \BibitemShut {NoStop}%
%%CITATION = ARXIV:1709.08649;%%
\bibitem [{\citenamefont {Fleischer}\ and\ \citenamefont
  {Jegerlehner}(1983)}]{Fleischer:1982af}%
  \BibitemOpen
  \bibfield  {author} {\bibinfo {author} {\bibfnamefont {J.}~\bibnamefont
  {Fleischer}}\ and\ \bibinfo {author} {\bibfnamefont {F.}~\bibnamefont
  {Jegerlehner}},\ }\href {\doibase 10.1016/0550-3213(83)90296-1} {\bibfield
  {journal} {\bibinfo  {journal} {Nucl. Phys.}\ }\textbf {\bibinfo {volume}
  {B216}},\ \bibinfo {pages} {469} (\bibinfo {year} {1983})}\BibitemShut
  {NoStop}%
%%CITATION = NUPHA,B216,469;%%
\bibitem [{\citenamefont {Denner}\ \emph {et~al.}(1992)\citenamefont {Denner},
  \citenamefont {Kublbeck}, \citenamefont {Mertig},\ and\ \citenamefont
  {Bohm}}]{Denner:1992bc}%
  \BibitemOpen
  \bibfield  {author} {\bibinfo {author} {\bibfnamefont {A.}~\bibnamefont
  {Denner}}, \bibinfo {author} {\bibfnamefont {J.}~\bibnamefont {Kublbeck}},
  \bibinfo {author} {\bibfnamefont {R.}~\bibnamefont {Mertig}}, \ and\ \bibinfo
  {author} {\bibfnamefont {M.}~\bibnamefont {Bohm}},\ }\href {\doibase
  10.1007/BF01555523} {\bibfield  {journal} {\bibinfo  {journal} {Z. Phys.}\
  }\textbf {\bibinfo {volume} {C56}},\ \bibinfo {pages} {261} (\bibinfo {year}
  {1992})}\BibitemShut {NoStop}%
%%CITATION = ZEPYA,C56,261;%%
\bibitem [{\citenamefont {Denner}\ \emph {et~al.}(2003)\citenamefont {Denner},
  \citenamefont {Dittmaier}, \citenamefont {Roth},\ and\ \citenamefont
  {Weber}}]{Denner:2003iy}%
  \BibitemOpen
  \bibfield  {author} {\bibinfo {author} {\bibfnamefont {A.}~\bibnamefont
  {Denner}}, \bibinfo {author} {\bibfnamefont {S.}~\bibnamefont {Dittmaier}},
  \bibinfo {author} {\bibfnamefont {M.}~\bibnamefont {Roth}}, \ and\ \bibinfo
  {author} {\bibfnamefont {M.~M.}\ \bibnamefont {Weber}},\ }\href {\doibase
  10.1016/S0550-3213(03)00269-4} {\bibfield  {journal} {\bibinfo  {journal}
  {Nucl. Phys.}\ }\textbf {\bibinfo {volume} {B660}},\ \bibinfo {pages} {289}
  (\bibinfo {year} {2003})},\ \Eprint {http://arxiv.org/abs/hep-ph/0302198}
  {arXiv:hep-ph/0302198 [hep-ph]} \BibitemShut {NoStop}%
%%CITATION = HEP-PH/0302198;%%
\bibitem [{\citenamefont {Belanger}\ \emph {et~al.}(2003)\citenamefont
  {Belanger}, \citenamefont {Boudjema}, \citenamefont {Fujimoto}, \citenamefont
  {Ishikawa}, \citenamefont {Kaneko}, \citenamefont {Kato},\ and\ \citenamefont
  {Shimizu}}]{Belanger:2002ik}%
  \BibitemOpen
  \bibfield  {author} {\bibinfo {author} {\bibfnamefont {G.}~\bibnamefont
  {Belanger}}, \bibinfo {author} {\bibfnamefont {F.}~\bibnamefont {Boudjema}},
  \bibinfo {author} {\bibfnamefont {J.}~\bibnamefont {Fujimoto}}, \bibinfo
  {author} {\bibfnamefont {T.}~\bibnamefont {Ishikawa}}, \bibinfo {author}
  {\bibfnamefont {T.}~\bibnamefont {Kaneko}}, \bibinfo {author} {\bibfnamefont
  {K.}~\bibnamefont {Kato}}, \ and\ \bibinfo {author} {\bibfnamefont
  {Y.}~\bibnamefont {Shimizu}},\ }\href {\doibase
  10.1016/S0370-2693(03)00339-3} {\bibfield  {journal} {\bibinfo  {journal}
  {Phys. Lett.}\ }\textbf {\bibinfo {volume} {B559}},\ \bibinfo {pages} {252}
  (\bibinfo {year} {2003})},\ \Eprint {http://arxiv.org/abs/hep-ph/0212261}
  {arXiv:hep-ph/0212261 [hep-ph]} \BibitemShut {NoStop}%
%%CITATION = HEP-PH/0212261;%%
\bibitem [{ATL(2019)}]{ATL-PHYS-PUB-2019-009}%
  \BibitemOpen
  \href {http://cds.cern.ch/record/2667570} {\emph {\bibinfo {title}
  {{Constraint of the Higgs boson self-coupling from Higgs boson differential
  production and decay measurements}}}},\ \bibinfo {type} {Tech. Rep.}\
  \bibinfo {number} {ATL-PHYS-PUB-2019-009}\ (\bibinfo  {institution} {CERN},\
  \bibinfo {address} {Geneva},\ \bibinfo {year} {2019})\BibitemShut {NoStop}%
\bibitem [{\citenamefont {collaboration}(2019)}]{ATLAS:2019pbo}%
  \BibitemOpen
  \bibfield  {author} {\bibinfo {author} {\bibfnamefont {T.~A.}\ \bibnamefont
  {collaboration}} (\bibinfo {collaboration} {ATLAS}),\ }\href@noop {} {\
  (\bibinfo {year} {2019})}\BibitemShut {NoStop}%
%%CITATION = ATLAS-CONF-2019-049;%%
\bibitem [{\citenamefont {Abelleira~Fernandez}\ \emph
  {et~al.}(2012)\citenamefont {Abelleira~Fernandez} \emph
  {et~al.}}]{AbelleiraFernandez:2012cc}%
  \BibitemOpen
  \bibfield  {author} {\bibinfo {author} {\bibfnamefont {J.~L.}\ \bibnamefont
  {Abelleira~Fernandez}} \emph {et~al.} (\bibinfo {collaboration} {LHeC Study
  Group}),\ }\href {\doibase 10.1088/0954-3899/39/7/075001} {\bibfield
  {journal} {\bibinfo  {journal} {J. Phys.}\ }\textbf {\bibinfo {volume}
  {G39}},\ \bibinfo {pages} {075001} (\bibinfo {year} {2012})},\ \Eprint
  {http://arxiv.org/abs/1206.2913} {arXiv:1206.2913 [physics.acc-ph]}
  \BibitemShut {NoStop}%
%%CITATION = ARXIV:1206.2913;%%
\bibitem [{\citenamefont {Han}\ and\ \citenamefont
  {Mellado}(2010)}]{Han:2009pe}%
  \BibitemOpen
  \bibfield  {author} {\bibinfo {author} {\bibfnamefont {T.}~\bibnamefont
  {Han}}\ and\ \bibinfo {author} {\bibfnamefont {B.}~\bibnamefont {Mellado}},\
  }\href {\doibase 10.1103/PhysRevD.82.016009} {\bibfield  {journal} {\bibinfo
  {journal} {Phys. Rev.}\ }\textbf {\bibinfo {volume} {D82}},\ \bibinfo {pages}
  {016009} (\bibinfo {year} {2010})},\ \Eprint {http://arxiv.org/abs/0909.2460}
  {arXiv:0909.2460 [hep-ph]} \BibitemShut {NoStop}%
%%CITATION = ARXIV:0909.2460;%%
\bibitem [{\citenamefont {Kumar}\ \emph {et~al.}(2017)\citenamefont {Kumar},
  \citenamefont {Ruan}, \citenamefont {Islam}, \citenamefont {Cornell},
  \citenamefont {Klein}, \citenamefont {Klein},\ and\ \citenamefont
  {Mellado}}]{Kumar:2015kca}%
  \BibitemOpen
  \bibfield  {author} {\bibinfo {author} {\bibfnamefont {M.}~\bibnamefont
  {Kumar}}, \bibinfo {author} {\bibfnamefont {X.}~\bibnamefont {Ruan}},
  \bibinfo {author} {\bibfnamefont {R.}~\bibnamefont {Islam}}, \bibinfo
  {author} {\bibfnamefont {A.~S.}\ \bibnamefont {Cornell}}, \bibinfo {author}
  {\bibfnamefont {M.}~\bibnamefont {Klein}}, \bibinfo {author} {\bibfnamefont
  {U.}~\bibnamefont {Klein}}, \ and\ \bibinfo {author} {\bibfnamefont
  {B.}~\bibnamefont {Mellado}},\ }\href {\doibase
  10.1016/j.physletb.2016.11.039} {\bibfield  {journal} {\bibinfo  {journal}
  {Phys. Lett.}\ }\textbf {\bibinfo {volume} {B764}},\ \bibinfo {pages} {247}
  (\bibinfo {year} {2017})},\ \Eprint {http://arxiv.org/abs/1509.04016}
  {arXiv:1509.04016 [hep-ph]} \BibitemShut {NoStop}%
%%CITATION = ARXIV:1509.04016;%%
\bibitem [{\citenamefont {Kumar}\ \emph {et~al.}(2015)\citenamefont {Kumar},
  \citenamefont {Ruan}, \citenamefont {Cornell}, \citenamefont {Islam},\ and\
  \citenamefont {Mellado}}]{Kumar:2015tua}%
  \BibitemOpen
  \bibfield  {author} {\bibinfo {author} {\bibfnamefont {M.}~\bibnamefont
  {Kumar}}, \bibinfo {author} {\bibfnamefont {X.}~\bibnamefont {Ruan}},
  \bibinfo {author} {\bibfnamefont {A.~S.}\ \bibnamefont {Cornell}}, \bibinfo
  {author} {\bibfnamefont {R.}~\bibnamefont {Islam}}, \ and\ \bibinfo {author}
  {\bibfnamefont {B.}~\bibnamefont {Mellado}},\ }\bibfield  {booktitle} {\emph
  {\bibinfo {booktitle} {{Proceedings, Workshop on Discovery Physics at the LHC
  (Kruger 2014): Kruger National Park, Mpumalanga, South Africa, December 1-6,
  2014}}},\ }\href {\doibase 10.1088/1742-6596/623/1/012017} {\bibfield
  {journal} {\bibinfo  {journal} {J. Phys. Conf. Ser.}\ }\textbf {\bibinfo
  {volume} {623}},\ \bibinfo {pages} {012017} (\bibinfo {year}
  {2015})}\BibitemShut {NoStop}%
%%CITATION = 00462,623,012017;%%
\bibitem [{\citenamefont {Hahn}(2001)}]{Hahn:2000kx}%
  \BibitemOpen
  \bibfield  {author} {\bibinfo {author} {\bibfnamefont {T.}~\bibnamefont
  {Hahn}},\ }\href {\doibase 10.1016/S0010-4655(01)00290-9} {\bibfield
  {journal} {\bibinfo  {journal} {Comput. Phys. Commun.}\ }\textbf {\bibinfo
  {volume} {140}},\ \bibinfo {pages} {418} (\bibinfo {year} {2001})},\ \Eprint
  {http://arxiv.org/abs/hep-ph/0012260} {arXiv:hep-ph/0012260 [hep-ph]}
  \BibitemShut {NoStop}%
%%CITATION = HEP-PH/0012260;%%
\bibitem [{\citenamefont {Hahn}\ and\ \citenamefont
  {Perez-Victoria}(1999)}]{Hahn:1998yk}%
  \BibitemOpen
  \bibfield  {author} {\bibinfo {author} {\bibfnamefont {T.}~\bibnamefont
  {Hahn}}\ and\ \bibinfo {author} {\bibfnamefont {M.}~\bibnamefont
  {Perez-Victoria}},\ }\href {\doibase 10.1016/S0010-4655(98)00173-8}
  {\bibfield  {journal} {\bibinfo  {journal} {Comput. Phys. Commun.}\ }\textbf
  {\bibinfo {volume} {118}},\ \bibinfo {pages} {153} (\bibinfo {year}
  {1999})},\ \Eprint {http://arxiv.org/abs/hep-ph/9807565}
  {arXiv:hep-ph/9807565 [hep-ph]} \BibitemShut {NoStop}%
%%CITATION = HEP-PH/9807565;%%
\bibitem [{\citenamefont {Denner}(1993)}]{Denner:1991kt}%
  \BibitemOpen
  \bibfield  {author} {\bibinfo {author} {\bibfnamefont {A.}~\bibnamefont
  {Denner}},\ }\href {\doibase 10.1002/prop.2190410402} {\bibfield  {journal}
  {\bibinfo  {journal} {Fortsch. Phys.}\ }\textbf {\bibinfo {volume} {41}},\
  \bibinfo {pages} {307} (\bibinfo {year} {1993})},\ \Eprint
  {http://arxiv.org/abs/0709.1075} {arXiv:0709.1075 [hep-ph]} \BibitemShut
  {NoStop}%
%%CITATION = ARXIV:0709.1075;%%
\bibitem [{\citenamefont {Mertig}\ \emph {et~al.}(1991)\citenamefont {Mertig},
  \citenamefont {Bohm},\ and\ \citenamefont {Denner}}]{Mertig:1990an}%
  \BibitemOpen
  \bibfield  {author} {\bibinfo {author} {\bibfnamefont {R.}~\bibnamefont
  {Mertig}}, \bibinfo {author} {\bibfnamefont {M.}~\bibnamefont {Bohm}}, \ and\
  \bibinfo {author} {\bibfnamefont {A.}~\bibnamefont {Denner}},\ }\href
  {\doibase 10.1016/0010-4655(91)90130-D} {\bibfield  {journal} {\bibinfo
  {journal} {Comput. Phys. Commun.}\ }\textbf {\bibinfo {volume} {64}},\
  \bibinfo {pages} {345} (\bibinfo {year} {1991})}\BibitemShut {NoStop}%
%%CITATION = CPHCB,64,345;%%
\bibitem [{\citenamefont {Shtabovenko}\ \emph {et~al.}(2016)\citenamefont
  {Shtabovenko}, \citenamefont {Mertig},\ and\ \citenamefont
  {Orellana}}]{Shtabovenko:2016sxi}%
  \BibitemOpen
  \bibfield  {author} {\bibinfo {author} {\bibfnamefont {V.}~\bibnamefont
  {Shtabovenko}}, \bibinfo {author} {\bibfnamefont {R.}~\bibnamefont {Mertig}},
  \ and\ \bibinfo {author} {\bibfnamefont {F.}~\bibnamefont {Orellana}},\
  }\href {\doibase 10.1016/j.cpc.2016.06.008} {\bibfield  {journal} {\bibinfo
  {journal} {Comput. Phys. Commun.}\ }\textbf {\bibinfo {volume} {207}},\
  \bibinfo {pages} {432} (\bibinfo {year} {2016})},\ \Eprint
  {http://arxiv.org/abs/1601.01167} {arXiv:1601.01167 [hep-ph]} \BibitemShut
  {NoStop}%
%%CITATION = ARXIV:1601.01167;%%
\bibitem [{\citenamefont {Hahn}(2005)}]{Hahn:2004fe}%
  \BibitemOpen
  \bibfield  {author} {\bibinfo {author} {\bibfnamefont {T.}~\bibnamefont
  {Hahn}},\ }\href {\doibase 10.1016/j.cpc.2005.01.010} {\bibfield  {journal}
  {\bibinfo  {journal} {Comput. Phys. Commun.}\ }\textbf {\bibinfo {volume}
  {168}},\ \bibinfo {pages} {78} (\bibinfo {year} {2005})},\ \Eprint
  {http://arxiv.org/abs/hep-ph/0404043} {arXiv:hep-ph/0404043 [hep-ph]}
  \BibitemShut {NoStop}%
%%CITATION = HEP-PH/0404043;%%
\bibitem [{\citenamefont {Schmidt}\ \emph {et~al.}(2016)\citenamefont
  {Schmidt}, \citenamefont {Pumplin}, \citenamefont {Stump},\ and\
  \citenamefont {Yuan}}]{Schmidt:2015zda}%
  \BibitemOpen
  \bibfield  {author} {\bibinfo {author} {\bibfnamefont {C.}~\bibnamefont
  {Schmidt}}, \bibinfo {author} {\bibfnamefont {J.}~\bibnamefont {Pumplin}},
  \bibinfo {author} {\bibfnamefont {D.}~\bibnamefont {Stump}}, \ and\ \bibinfo
  {author} {\bibfnamefont {C.~P.}\ \bibnamefont {Yuan}},\ }\href {\doibase
  10.1103/PhysRevD.93.114015} {\bibfield  {journal} {\bibinfo  {journal} {Phys.
  Rev.}\ }\textbf {\bibinfo {volume} {D93}},\ \bibinfo {pages} {114015}
  (\bibinfo {year} {2016})},\ \Eprint {http://arxiv.org/abs/1509.02905}
  {arXiv:1509.02905 [hep-ph]} \BibitemShut {NoStop}%
%%CITATION = ARXIV:1509.02905;%%
\bibitem [{\citenamefont {Alwall}\ \emph {et~al.}(2014)\citenamefont {Alwall},
  \citenamefont {Frederix}, \citenamefont {Frixione}, \citenamefont {Hirschi},
  \citenamefont {Maltoni} \emph {et~al.}}]{Alwall:2014hca}%
  \BibitemOpen
  \bibfield  {author} {\bibinfo {author} {\bibfnamefont {J.}~\bibnamefont
  {Alwall}}, \bibinfo {author} {\bibfnamefont {R.}~\bibnamefont {Frederix}},
  \bibinfo {author} {\bibfnamefont {S.}~\bibnamefont {Frixione}}, \bibinfo
  {author} {\bibfnamefont {V.}~\bibnamefont {Hirschi}}, \bibinfo {author}
  {\bibfnamefont {F.}~\bibnamefont {Maltoni}},  \emph {et~al.},\ }\href
  {\doibase 10.1007/JHEP07(2014)079} {\bibfield  {journal} {\bibinfo  {journal}
  {JHEP}\ }\textbf {\bibinfo {volume} {1407}},\ \bibinfo {pages} {079}
  (\bibinfo {year} {2014})},\ \Eprint {http://arxiv.org/abs/1405.0301}
  {arXiv:1405.0301 [hep-ph]} \BibitemShut {NoStop}%
%%CITATION = ARXIV:1405.0301;%%
\bibitem [{\citenamefont {Tanabashi}\ \emph {et~al.}(2018)\citenamefont
  {Tanabashi} \emph {et~al.}}]{Tanabashi:2018oca}%
  \BibitemOpen
  \bibfield  {author} {\bibinfo {author} {\bibfnamefont {M.}~\bibnamefont
  {Tanabashi}} \emph {et~al.} (\bibinfo {collaboration} {Particle Data
  Group}),\ }\href {\doibase 10.1103/PhysRevD.98.030001} {\bibfield  {journal}
  {\bibinfo  {journal} {Phys. Rev.}\ }\textbf {\bibinfo {volume} {D98}},\
  \bibinfo {pages} {030001} (\bibinfo {year} {2018})}\BibitemShut {NoStop}%
%%CITATION = PHRVA,D98,030001;%%
\bibitem [{\citenamefont {Klein}(2019)}]{Klein:2018rhq}%
  \BibitemOpen
  \bibfield  {author} {\bibinfo {author} {\bibfnamefont {M.}~\bibnamefont
  {Klein}},\ }in\ \href {\doibase 10.1142/9789813238053_0015} {\emph {\bibinfo
  {booktitle} {From My Vast Repertoire ...: Guido Altarelli's Legacy}}},\
  \bibinfo {editor} {edited by\ \bibinfo {editor} {\bibfnamefont
  {A.}~\bibnamefont {Levy}}, \bibinfo {editor} {\bibfnamefont {S.}~\bibnamefont
  {Forte}}, \ and\ \bibinfo {editor} {\bibfnamefont {G.}~\bibnamefont
  {Ridolfi}}}\ (\bibinfo {year} {2019})\ pp.\ \bibinfo {pages} {303--347},\
  \Eprint {http://arxiv.org/abs/1802.04317} {arXiv:1802.04317 [hep-ph]}
  \BibitemShut {NoStop}%
%%CITATION = ARXIV:1802.04317;%%
\bibitem [{\citenamefont {Abada}\ \emph {et~al.}(2019)\citenamefont {Abada}
  \emph {et~al.}}]{Abada:2019lih}%
  \BibitemOpen
  \bibfield  {author} {\bibinfo {author} {\bibfnamefont {A.}~\bibnamefont
  {Abada}} \emph {et~al.} (\bibinfo {collaboration} {FCC}),\ }\href {\doibase
  10.1140/epjc/s10052-019-6904-3} {\bibfield  {journal} {\bibinfo  {journal}
  {Eur. Phys. J.}\ }\textbf {\bibinfo {volume} {C79}},\ \bibinfo {pages} {474}
  (\bibinfo {year} {2019})}\BibitemShut {NoStop}%
%%CITATION = EPHJA,C79,474;%%
\bibitem [{\citenamefont {Mashiro}()}]{Mashiro:2017aqd}%
  \BibitemOpen
  \bibfield  {author} {\bibinfo {author} {\bibfnamefont {T.}~\bibnamefont
  {Mashiro}},\ }\bibfield  {booktitle} {\emph {\bibinfo {booktitle}
  {{Proceedings, 25th International Workshop on Deep-Inelastic Scattering and
  Related Topics (DIS 2017): Birmingham, UK, April 3-7, 2017}}},\ }\href
  {https://indico.cern.ch/event/568360/contributions/2523555/attachments/1440097/2216668/mtanaka_DIS2017_SMhiggs_final.pdf}
  {\ }\BibitemShut {NoStop}%
\bibitem [{\citenamefont {Jager}(2010)}]{Jager:2010zm}%
  \BibitemOpen
  \bibfield  {author} {\bibinfo {author} {\bibfnamefont {B.}~\bibnamefont
  {Jager}},\ }\href {\doibase 10.1103/PhysRevD.81.054018} {\bibfield  {journal}
  {\bibinfo  {journal} {Phys. Rev.}\ }\textbf {\bibinfo {volume} {D81}},\
  \bibinfo {pages} {054018} (\bibinfo {year} {2010})},\ \Eprint
  {http://arxiv.org/abs/1001.3789} {arXiv:1001.3789 [hep-ph]} \BibitemShut
  {NoStop}%
%%CITATION = ARXIV:1001.3789;%%
\end{thebibliography}
%merlin.mbs apsrev4-1.bst 2010-07-25 4.21a (PWD, AO, DPC) hacked
%Control: key (0)
%Control: author (72) initials jnrlst
%Control: editor formatted (1) identically to author
%Control: production of article title (-1) disabled
%Control: page (0) single
%Control: year (1) truncated
%Control: production of eprint (0) enabled
%

\end{document}